\documentclass[
aps,10pt, longbibliography, superscriptaddress,amsmath, amssymb, reprint
]{revtex4-2}

\usepackage[english]{babel}
\usepackage[T1]{fontenc}
\usepackage{amsmath,amssymb,bbm,mathrsfs,bm,braket,color,graphicx,comment,amsfonts,dsfont}
\usepackage{siunitx,physics}
\usepackage[colorlinks,citecolor=blue,urlcolor=blue]{hyperref}
\usepackage[version=4]{mhchem}
\usepackage[normalem]{ulem}
\usepackage{makecell}

\begin{document}

\title{Readout sweet spots for spin qubits with strong spin-orbit interaction}

\author{Domonkos Svastits}
\affiliation{Department of Theoretical Physics, Institute of Physics, Budapest University of Technology and Economics, Műegyetem rkp. 3., H-1111 Budapest, Hungary}
\affiliation{Qutility @ Faulhorn Labs, Budapest, Hungary}

\author{Bence Hetényi}
\affiliation{IBM Research Europe – Zurich, Säumerstrasse 4, 8803 Rüschlikon, Switzerland}

\author{Gábor Széchenyi}
\affiliation{ELTE Eötvös Loránd University, Institute of Physics, H-1117 Budapest, Hungary}

\author{James Wootton}
\affiliation{IBM Research Europe – Zurich, Säumerstrasse 4, 8803 Rüschlikon, Switzerland}
\affiliation{Moth Quantum AG, Schorenweg 44B, 4144, Arlesheim, Switzerland}

\author{Daniel Loss}
\affiliation{Department of Physics, University of Basel, Klingelbergstrasse 82, CH-4056 Basel, Switzerland}

\author{Stefano Bosco}
\affiliation{QuTech and Kavli  Institute of Nanoscience, Delft University of Technology, Delft, The Netherlands}

\author{András Pályi}
\affiliation{Department of Theoretical Physics, Institute of Physics, Budapest University of Technology and Economics, Műegyetem rkp. 3., H-1111 Budapest, Hungary}
\affiliation{HUN-REN-BME-BCE Quantum Technology Research Group, Műegyetem rkp. 3., H-1111 Budapest, Hungary}
\date{\today}

\begin{abstract}
Qubit readout schemes often deviate from ideal projective measurements, introducing critical issues that limit quantum computing performance. 
In this work, we model charge-sensing-based readout for semiconductor spin qubits in double quantum dots, and identify key error mechanisms caused by the back-action of the charge sensor.
We quantify how the charge noise of the sensor, residual tunneling, and $g$-tensor modulation degrade readout fidelity, induce a mixed post-measurement state, and cause leakage from the computational subspace. 
For state-of-the-art systems with strong spin-orbit interaction and electrically tunable $g$-tensors, we identify a readout sweet spot, that is, a special device configuration where readout is closest to projective.
Our framework provides a foundation for developing effective readout error mitigation strategies, with broad applications for optimizing readout performance for a variety of charge-sensing techniques, advancing quantum protocols, and improving adaptive circuits for error correction. 
\end{abstract}

\maketitle

\emph{Introduction.}
Spin qubits confined in semiconductor quantum dots \cite{Loss_1998} are front-runners for large-scale quantum computers \cite{HansonRMP, ZwanenburgRMP, eriksson, Burkard_2023, doi:10.1126/science.ado5915, Zhang2025, Philips2022, Hendrickx2021}. Their small size, scalability \cite{Burkard_2023, loss-prospects, Vandersypen2017, Scappucci2021, Stano2022}, tunability, CMOS compatibility \cite{Maurand2016, Xue2021, Jirovec2021, Liles2024, Camenzind2022, Zwerver2022, Neyens2024, Geyer2024, steinacker2024300mmfoundrysilicon}, and the demonstrated high-fidelity operations \cite{Xue2022, Noiri2022, doi:10.1126/sciadv.abn5130} and readout \cite{Takeda2024}
make them particularly attractive.

A key component of quantum computers is readout, an essential benchmark of which is readout fidelity. 
However,  while experiments demonstrate increasingly high fidelities \cite{PhysRevX.8.021046, PRXQuantum.3.010352, PRXQuantum.4.010329, steinacker2024300mmfoundrysilicon, Vigneau_2023}, 
this is not the only relevant benchmark \cite{Pereira_2022, Pereira_2023}.
Improving other metrics of mid-circuit measurements \cite{rudinger2022, ZhangZhihan2025, Hines2025, hothem2024measuringerrorratesmidcircuit, mclaren2025benchmarkingquantuminstruments},
such as the quality of the post-measurement state, and readout-induced leakage  \cite{Miao2023, McEwen2021, Battisel2021, Marques2023},
are key requirements for quantum error correction \cite{Liepelt_2024,Geher2025} and other dynamic circuits.
\cite{PhysRevApplied.17.064061, PRXQuantum.5.030339, Fleury_2024, stenger2025methodsimulatingopensystemdynamics, bäumer2024quantumfouriertransformusing, cao2025measurementdrivenquantumadvantagesshallow}

We present ways to simultaneously improve different metrics of state-of-the-art spin qubit readout. Our model focuses on  widely-used  measurements via Pauli blockade spin-to-charge conversion \cite{ono2002current, dots5coherent, koppens2006driven} and charge sensing in a double quantum dot (DQD) \cite{PhysRevLett.134.023601, Taubert_2008, ferguson2020quantummeasurementinducesmanybody, young2010inelastic}. By modeling charge sensing via a DC-biased quantum point contact (QPC) \cite{ElzermanChargeSensing,Elzerman,Gustavsson, PhysRevLett.75.3340, PhysRevB.93.195411}, we show that high-fidelity and back-action-free readout requires switching off the inter-dot tunnel coupling upon charge sensing.

In current devices with strong spin-orbit interaction (e.g., holes in Ge and Si), the fluctuating electric field induced by the carriers of the QPC modulates the g-tensors of the spins in the DQD.
We show that this modulation influences the readout quality, depending on the static homogeneous magnetic field applied to the DQD. 
We identify a readout sweet spot, i.e., a specific device configuration where the detrimental back-action effects of the g-tensor modulation are minimized: leakage is suppressed and the purity of the post-measurement state is maintained.
We argue that this readout sweet spot exists universally, independent of spin-orbit interaction details.
Our findings provide practical guidelines for high-quality spin-qubit readout.

\emph{Model.}
To describe the readout of a qubit defined in a DQD, we rely on the general framework of quantum measurements, where the \textit{system} is measured indirectly through a \textit{meter}  \cite{Wiseman_Milburn_2009, diosi2007short}.
Simplifying earlier approaches \cite{Goan_2001_1, Goan_2001_2}, we use a model
in which the meter (the QPC) is represented by a qubit, hence we call it the \textit{Qubit Measures Qubit (QMQ) model}.
To illustrate the harmful effect of residual tunneling in the DQD, we use the model to describe charge qubit readout.
Then, we generalize the QMQ model to the readout of spin qubits, 
where a residual tunneling has similar negative effects. We compare the QMQ model to earlier approaches\cite{Goan_2001_1, Goan_2001_2, Landauer, PhysRevLett.57.1761, BLANTER2001} in SM~S7\cite{SuppMat}.

\begin{figure}
    \centering
    \includegraphics[width=\linewidth]{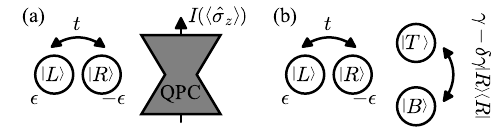}
    \caption{Charge-sensing-based readout of a charge qubit.
    (a) A double-dot charge qubit is read out by a quantum point contact (QPC). 
    (b) QPC current is modeled as a sequence of indirect measurements via another qubit (the meter), with basis states $\ket{B}$ and $\ket{T}$.
    The meter is Coulomb-coupled to dot R of the charge qubit with interaction strength $\delta \gamma$, 
    hence the probability of tunneling through the meter depends on the charge qubit state.
    }
    \label{fig:qpc-scheme}
\end{figure}

We consider the setup shown in Fig.~\ref{fig:qpc-scheme}a.
The charge qubit is formed by a single carrier in the DQD; the qubit basis states $\ket{L}$ and $\ket{R}$ correspond to its position.
The QPC current is high (low) if the DQD charge is in the left (right) dot, due to Coulomb repulsion between the DQD charge and QPC electrons.
In our QMQ model, the QPC is also represented by a single-carrier DQD,
see Fig.~\ref{fig:qpc-scheme}b, with basis states corresponding to the bottom ($\ket{B}$) and top ($\ket{T}$) dots.

The Hamiltonian of the qubit-meter system reads
\begin{align}
    \hat{H}_\mathrm{tot} &= \hat{H}_\mathrm{charge} + \hat{H}_\mathrm{m} + \hat{H}_\mathrm{int}^\mathrm{c},
    \label{eq:charge-hamiltonian}
\end{align}
where $\hat{H}_\mathrm{charge} = \epsilon \hat{\sigma}_z + t \hat{\sigma}_x$, $\hat{H}_\mathrm{m} = \gamma \hat{\tau}_x$, $\hat{H}_\mathrm{int}^\mathrm{c} = -\delta \gamma \ket{R}\bra{R}  \hat{\tau}_x$, $\hat{\sigma}_z = \ket{L}\bra{L}-\ket{R}\bra{R}$, $\hat{\sigma}_x = \ket{L}\bra{R}+\ket{R}\bra{L}$ and $\hat{\tau}_x = \ket{T}\bra{B}+\ket{B}\bra{T}$. 
Furthermore, $2\epsilon$ is the on-site energy detuning, $t$ is the tunneling amplitude 
of the qubit, $\gamma$ is the tunneling amplitude 
of the meter, and the interaction strength $\delta \gamma$ describes the Coulomb repulsion between the qubit and the meter.

One way to operate the charge qubit \cite{PeterssonPRL, ScarlinoPRX} is to perform coherent control at $t>0$ and $\epsilon = 0$, and sweep $\epsilon$ and $t$ to a readout point with $\epsilon \gg t$ \cite{meinersen2024}. Then, readout in the basis of the ground $\ket{\mathrm{g}}$ and excited $\ket{\mathrm{e}}$ states (at the readout point) is carried out by measuring the QPC current and inferring binary value g or e from the current trace. 

The QMQ model describes the process of reading out the qubit state $\rho_\mathrm{pre}$. The readout is modeled as a sequence of indirect measurements, each consisting of three steps:
(1) The meter is initialized in state $\ket{B}$. 
This state represents an electron approaching the QPC from the bottom lead. 
(2) The qubit-meter system evolves unitarily according to $\hat{H}_\mathrm{tot}$ for time $\Delta \tau$, entangling data and meter qubits via the interaction $\hat{H}_\mathrm{int}^\mathrm{c}$.
The meter's tunneling rate $\gamma$ and the timestep $\Delta \tau$ are set such that by time $\Delta \tau$, the transition probability to state $\ket{T}$ is $1/2$ in an equal superposition or mixture of $\ket{L}$ and $\ket{R}$. This corresponds to a QPC tuned to its charge-sensing working point where its conductance is $1/2$ of the conductance quantum.
(3) The meter is measured projectively in the $\{\ket{B}, \ket{T}\}$ basis. The assigned measurement outcome is the number of transmitted electrons, i.e., 0 (1) if the QPC electron is found in the bottom (top) dot. 

The number $N\gg1$ of these indirect measurements corresponds to the number of electrons attempting to transit the QPC during readout. 
The current flowing through the QPC is represented as an $N$-long bitstring, which contains 1-s where the QPC electron was transmitted.
The state of the qubit is inferred from the 
transmission ratio $k = N_\mathrm{t}/N$ using the maximum-likelihood principle, where $N_\mathrm{t}$ is the number of electrons transmitted.
That is, we infer e if the probability of obtaining the measured transmission ratio
is greater for the initial state $\ket{\mathrm{e}}$ then for $\ket{\mathrm{g}}$; we infer g otherwise.

\emph{Measurement and back-action.}
The quantum channels corresponding to the final outcomes e and g are represented by superoperators $\mathcal{M}_{\mathrm{e}}$ and $\mathcal{M}_{\mathrm{g}}$.
They are called \emph{measurement operations} \cite{Wiseman_Milburn_2009, hashim2024practicalintroductionbenchmarkingcharacterization}, and they map the pre-measurement state to the unnormalized post-measurement state conditional on the outcome.
In the QMQ model, these measurement operations can be efficiently computed numerically (see SM~S1~E \cite{SuppMat}). 
Since $\mathcal{M}_{\mathrm{g}}$ and $\mathcal{M}_{\mathrm{e}}$  completely characterize the measurement, we extract all relevant readout metrics from them.

\begin{figure}
    \centering
    \includegraphics[width=\linewidth]{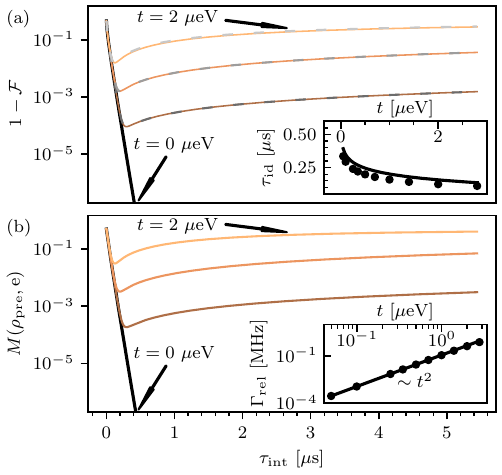}
    \caption{Measurement benchmarks for charge qubit readout. (a) Infidelity. The continuous lines show the numerical values of infidelity.
    The dashed lines show an analytical estimate of the infidelity (see SM~S5\cite{SuppMat}, which relies on Ref.~\onlinecite{spacings}). 
    Inset: Optimal integration time determined from exact numerical calculations (dots) 
    and analytical estimate (line) (see SM~S5\cite{SuppMat}). 
    (b) Mixedness calculated numerically for $\rho_\mathrm{pre} = (\ket{\mathrm{e}}\bra{\mathrm{e}}+\ket{\mathrm{g}}\bra{\mathrm{g}})/2$ and outcome e. Inset: Relaxation rate from numerical (dots) and analytical (line) calculations. We used the following parameters for the plots: $\epsilon=10 \ \mathrm{\mu eV}$, $\gamma = 5 \ \mathrm{\mu eV}$, $\delta\gamma = 0.5 \ \mathrm{\mu eV}$, $\Delta \tau = 0.11 \ \mathrm{ns}$.
    The latter three values correspond to experimentally realistic current values (SM~S2\cite{SuppMat}).
    The values of the residual tunneling used are $t\in\{0, 0.1, 0.5, 2\} \ \mathrm{\mu eV}$.}
    \label{fig:quantities-charge}
\end{figure}

Infidelity is expressed with the measurement operations as
\begin{equation}
    1-\mathcal{F} \equiv 1 - \frac{1}{2} \left(\Tr{\mathcal{M}_{\mathrm{e}}[\ket{\mathrm{e}}\bra{\mathrm{e}}]}+\Tr{\mathcal{M}_{\mathrm{g}}[\ket{\mathrm{g}}\bra{\mathrm{g}}]}\right).\label{eq:fidel-def}
\end{equation}
Fig.~\ref{fig:quantities-charge}a shows the infidelity as a function of integration time $\tau_\mathrm{int} = N \Delta \tau$.
For short integration times, the infidelity decreases with increasing $\tau_\mathrm{int}$,
since we gain more information about the initial state. 
For $t=0$, the infidelity vanishes as $\tau_\mathrm{int} \rightarrow \infty$.
However, for $t\neq 0$, the infidelity increases after an ideal integration time $\tau_\mathrm{id}$. 
As argued below, this increase is due to measurement back-action leading to qubit relaxation.

The time dependence of the infidelity in Fig.~\ref{fig:quantities-charge}a is explained as follows.
The pace of information gain upon readout is quantified by the \emph{measurement rate} $\Gamma_\mathrm{m}$, i.e., the decay rate of the overlap between the two distributions of the transmission ratio $k$ corresponding to pre-measurement states $\ket{\mathrm{g}}$ and $\ket{\mathrm{e}}$. This overlap has the form $1-\Phi(\sqrt{2\Gamma_\mathrm{m}\tau_\mathrm{int}})$, through which $\Gamma_\mathrm{m}$ is defined, where $\Phi$ is the cumulative distribution function of the normal distribution. 
Upon calculating $\Gamma_\mathrm{m}$ up to lowest (zeroth) order in $t/\epsilon$ (see SM~S3\cite{SuppMat}), which is the relevant limit 
at the readout point, we find 
\begin{equation}
\label{eq:measurementrate}
    \Gamma_\mathrm{m} = \frac{1}{2}\left(\frac{\delta\gamma}{\hbar}\right)^2\Delta\tau.
\end{equation}
This rate characterizes the decrease of the infidelity in Fig.~\ref{fig:quantities-charge}a for short integration times.

A nonzero residual tunneling $t>0$ implies 
$[\hat{H}_\mathrm{charge}, \hat{H}_\mathrm{int}^\mathrm{c}]\neq 0$.
This, in turn, implies readout back-action in the form of qubit relaxation \cite{Goan_2001_1, Boissonneault_2009} characterized by the \emph{relaxation rate} $\Gamma^\mathrm{c}_\mathrm{rel}$, which we express (see SM~S4\cite{SuppMat}) as:
\begin{equation}
    \Gamma_\mathrm{rel}^\mathrm{c} = \frac{1}{2}\frac{t^2\delta\gamma^2}{\epsilon^4 \Delta\tau}\sin^2\left(\frac{\epsilon\Delta\tau}{\hbar}\right).
    \label{eq:rel-rate}
\end{equation} 
This rate quantifies the speed at which the unconditional post-measurement state $\mathcal{M}[\rho_\mathrm{in}] = \mathcal{M}_{\mathrm{g}}[\rho_\mathrm{in}] + \mathcal{M}_{\mathrm{e}}[\rho_\mathrm{in}]$ approaches the completely mixed state. 
The inset of Fig.~\ref{fig:quantities-charge}b shows that Eq.~\eqref{eq:rel-rate} (solid line) matches the relaxation rates obtained numerically (points) for the investigated parameter range. 
Note that Eq.~\eqref{eq:rel-rate} is not valid if the argument of the sine is close to $n\pi$, $n\in \mathbb{Z}^+$ (see SM~S4\cite{SuppMat}).

The second readout metric we consider is the mixedness of the post-measurement state. 
In a projective measurement of a qubit, 
the post-measurement state is pure.
However, in our setup, the post-measurement state is mixed, except for the special case $t=0$ and $\tau_\mathrm{int} \to \infty$.
The post-measurement state depends on the pre-measurement state $\rho_\mathrm{pre}$ and the measurement outcome $r\in\{\mathrm{e,g}\}$. 
We express the post-measurement state mixedness with the measurement operations as
\begin{equation}
    M(\rho_\mathrm{pre}, r) \equiv 1 -\Tr{\rho_{\mathrm{post}, r}^2},
\end{equation}
with the conditional post-measurement state $\rho_\mathrm{post, r} = \mathcal{M}_r[\rho_\mathrm{pre}]/\Tr{\mathcal{M}_r[\rho_\mathrm{pre}]}$.
A nonzero mixedness indicates deviations from a projective measurement. 
In Fig.~\ref{fig:quantities-charge}b, the mixedness for outcome $r=\mathrm{e}$ is shown for the fully mixed pre-measurement state, as a function of integration time.
The qualitative dependence of the mixedness on integration time is similar to that of the infidelity, and is well described by the measurement and relaxation rates.

These results confirm that a residual tunneling degrades readout, making precise control of $t$ essential.
Below, we argue that the same conclusion applies for Pauli-blockade-based spin qubit readout.

\begin{figure*}
    \centering
    \includegraphics[width=\linewidth]{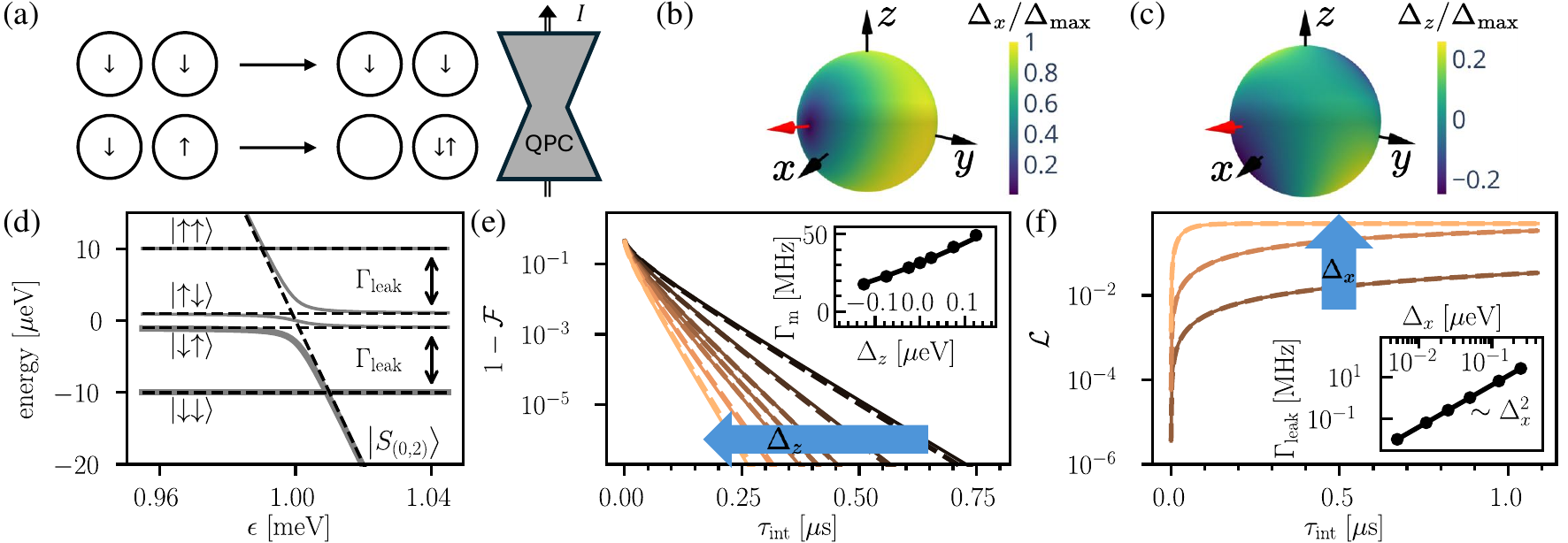}
    \caption{
    Spin-qubit readout error due to $g$-tensor modulation by the charge sensor.
    (a) Readout setup in a double quantum dot: 
    Spin in the right dot is to be read out utilizing the reference spin in the left dot, with the combination of Pauli spin blockade and charge sensing.
    (b), (c) Readout sweet spot, shown by red arrow: magnetic field direction providing optimized readout ($\Delta_x = 0$), for $g$-tensor parameters from Ref.~\onlinecite{Crippa_2018}. 
    Colors show normalized $\Delta_x$ (b)  and $\Delta_z$ (c) as functions of the magnetic field direction. The normalization factor $\Delta_\mathrm{max}$ is the maximum of $\Delta_x$ as a function of magnetic field direction.
    (d) Lowest five energy levels of the double dot as a function of detuning for $t=0$ (dashed black lines) and $t=2 \ \mathrm{\mu eV}$ (grey solid lines), for $U=1$ meV, $Z_L=11 \ \mathrm{\mu eV}$, and
    $Z_R=9 \ \mathrm{\mu eV}$. 
    Arrows indicate incoherent QPC-induced transitions activated by a nonzero $\Delta_x$. 
    (e) Readout infidelity for $\Delta_x = 0$ and 
    $\Delta_z \in 
    \{-0.125,
    -0.075,
    -0.025,
    0,
    0.025,
    0.075,
    0.125\} \ \mu \mathrm{eV}$.
    Arrow indicates increasing $\Delta_z$. 
    Inset: Analytical (solid) and numerical (dots) results for the measurement rate $\Gamma_\mathrm{m}$ as function of $\Delta_z$. 
    (f) Leakage from the initial state $\ket{\downarrow\downarrow}$ for $\Delta_z = 0$ and for 
    $\Delta_x \in \{0.0125, 0.05, 0.25\}\, \mu\mathrm{eV}$. 
    Arrow indicates increasing $\Delta_x$. 
    Inset: Analytical (solid) and numerical (dots) results for the relaxation rate $\Gamma_\mathrm{leak}$ as a function of $\Delta_x$.
    Parameters for (e, f): 
    $\epsilon=U+40 \ \mathrm{\mu eV}$, $t = 0 \ \mathrm{\mu eV}$, $\gamma = 5 \ \mathrm{\mu eV}$, $\delta\gamma = 0.5 \ \mathrm{\mu eV}$, 
    $U$, $Z_L$, $Z_R$ as above.}
    \label{fig:spin-setup}
\end{figure*}

\emph{Spin qubit.} Spin qubit readout often relies on Pauli blockade spin-to-charge conversion and subsequent charge sensing \cite{PRXQuantum.4.010329, PRXQuantum.3.010352, PhysRevX.8.021046}. 
We generalize the QMQ model to describe this readout.
We focus on Pauli-blockade readout of a single-electron (Loss-DiVincenzo \cite{Loss_1998}) spin qubit. 
This requires a DQD with a target spin to be measured in, say, the right dot and a known reference spin in the left dot, see Fig.~\ref{fig:spin-setup}a.
Readout is done in three steps: spin-to-charge conversion, 
charge sensing and charge-to-spin conversion. 
Without relaxation processes, 
this readout is non-destructive.

We model the two-electron DQD by the Hamiltonian of a two-site Hubbard model:
\begin{equation}
    \hat{H}_\mathrm{spin} = \hat{H}_\mathrm{on-site}+ \hat{H}_\mathrm{tun}+\hat{H}_\mathrm{Zeeman}+\hat{H}_\mathrm{Coulomb},
    \label{eq:hubbard}
\end{equation}
where $\hat{H}_\mathrm{on-site} = \epsilon/2 \sum_s \left(\hat{n}_{L, s}-\hat{n}_{R,s}\right)$, $\hat{H}_\mathrm{tun} =t\sum_{s}\left(\hat{a}_{R, s}^\dagger\hat{a}_{L,s}+h.c.\right)$, $\hat{H}_\mathrm{Zeeman} = Z_L/2 \left(\hat{n}_{L\uparrow}-\hat{n}_{L\downarrow}\right) + Z_R/2 \left(\hat{n}_{R\uparrow}-\hat{n}_{R\downarrow}\right)$, $\hat{H}_\mathrm{Coulomb} = U\left(\hat{n}_{L,\uparrow}\hat{n}_{L,\downarrow}+\hat{n}_{R,\uparrow}\hat{n}_{R,\downarrow}\right)$,  $\hat{a}^\dag_{\sigma,s}$ is an electronic creation operator on dot $\sigma \in \{L,R\}$ with spin $s\in\{\uparrow,\downarrow\}$, and  $\hat{n}_{\sigma, s} = \hat{a}^\dag_{\sigma, s} \hat{a}_{\sigma,s}$. Furthermore, $Z_{\sigma} = \mu_B g_{\sigma}B_z$ is the Zeeman energy in dot $\sigma$ due to a magnetic field $B$, proportional to the $g$-factor $g_\sigma$ of dot $\sigma$, and $U$ is the  Coulomb repulsion energy between two electrons occupying the same dot. 
At this point, spin-orbit interaction is included in Eq.~\eqref{eq:hubbard} by considering different $g$-factors in the two dots.

For later use, we introduce the six two-electron basis states
$\ket{S_{(0,2)}} = \hat{a}^\dag_{R\uparrow} \hat{a}^\dag_{R\downarrow} \ket{0}$,
$\ket{\uparrow \downarrow} = 
\hat{a}^\dag_{L\uparrow}
\hat{a}^\dag_{R\downarrow}
\ket{0}$, etc,
with $\ket{0}$ denoting the state of the empty DQD.
These states are eigenstates of $\hat{H}_\mathrm{spin}$ at $t=0$.
The detuning dependence of the five lowest-energy states for $t=0$ is illustrated in Fig.~\ref{fig:spin-setup}d as the dashed lines. We focus on readout errors due to the back-action of the charge sensor. Therefore, we assume perfect state transfer during spin-to-charge and charge-to-spin conversion between the computational states at the operational point (left end of Fig.~\ref{fig:spin-setup}d) and the the computational states at the readout point (right end of Fig.~\ref{fig:spin-setup}d) according to, say, $\ket{\downarrow \uparrow} \leftrightarrow \ket{S_{0,2}}$ and $\ket{\downarrow \downarrow} \leftrightarrow \ket{\downarrow\downarrow}$.

The interaction between the DQD and the QPC is described by the term 
\begin{equation}
    \hat{H}_\mathrm{int}^\mathrm{s} = -\delta\gamma \ket{S_{(0, 2)}}\bra{S_{(0, 2)}}\otimes \hat{\tau}_x,
    \label{eq:int-hamiltonian-spin0}
\end{equation}
which is a natural generalization of $\hat{H}_\mathrm{int}^\mathrm{c}$ in Eq.~\eqref{eq:charge-hamiltonian}.
This $\hat{H}_\mathrm{int}^\mathrm{s}$ captures that the transmission probability of the QPC is lower if there are two electrons in the right dot, instead of one.

We use this model to describe readout of the spin qubit in dot $R$. The measurement operations $\mathcal{M}_\downarrow$ and $\mathcal{M}_\uparrow$ are computed analogously to those of the charge qubit readout as a function of model parameters.

Ideally, tunnel coupling is switched off at the readout position during charge sensing.
State-of-the-art device fabrication makes this possible \cite{PRXQuantum.4.010329}, and shuttling in a sparse dot array \cite{doi:10.1126/science.ado5915} provides an alternative solution.
However, in a given device, it might be impossible to set $t=0$, e.g., due to the gate electrode geometry, or gate-voltage constraints.
This enhances readout errors, akin to the case of the charge qubit discussed above. 
A small but nonzero residual tunneling fulfilling $t \ll \epsilon-U$ and $4t^2/(\epsilon-U) \ll |Z_L - Z_R|$ implies that $\ket{S_{(0,2)}}$
is slightly hybridized
with $\ket{\uparrow \downarrow}$ and $\ket{\downarrow\uparrow}$; the perturbed eigenstates are denoted by 
$\ket{\Tilde{S}_{(0,2)}}$,
$\ket{\Tilde{\uparrow\downarrow}}$, and
$\ket{\Tilde{\downarrow\uparrow}}$. 
Consequently, the interaction of the DQD with the QPC electrons leads to incoherent transitions between the state $\ket{\Tilde{S}_{(0,2)}}$ and the states $\ket{\Tilde{\uparrow\downarrow}}$ and $\ket{\Tilde{\downarrow\uparrow}}$.
The measurement benchmarks infidelity and mixedness show the same qualitative behavior as illustrated in Fig.~\ref{fig:quantities-charge} for the charge qubit. 
This back-action is discussed further in SM~S6\cite{SuppMat}.

From now on, we assume that the tunnel coupling can be switched off ($t=0$).
In this case, another mechanism affecting readout is the spin-orbit mediated interaction between the DQD spins and the fluctuating electric field created by the QPC electrons \cite{KatoScience,Borhani,Crippa_2018, Venitucci_2018,Studenikin19}, often described as $g$-tensor modulation.
To account for this, we generalize the interaction Hamiltonian of Eq.~\eqref{eq:int-hamiltonian-spin0} as
\begin{equation}
    \hat{H}_\mathrm{int}^\mathrm{s} = -\left(\delta\gamma\ket{S_{(0,2)}}\bra{S_{(0,2)}} + \mathbf{\hat{s}}_R
    \cdot
    \mathbf{\Delta} \right)\otimes \hat{\tau}_x,
\label{eq:int-hamiltonian-spin}
\end{equation}
where $\mathbf{\Delta} = \mu_\mathrm{B} g'_R \mathbf{B}/2$ describes the \emph{spin-charge coupling}, i.e., 
change of the $g$-tensor on dot $R$ caused by the electric field of the charges flowing through the QPC. 
Here, $g'_R$ is a $3\times 3$ real matrix,
and 
$\mathbf{\hat{s}}_R$ represents the spin vector on the right dot, e.g., $\hat{s}_{R,z} = \hat{n}_{R\uparrow} - \hat{n}_{R\downarrow}$.
The joint Hamiltonian of the system and the meter is given by Eq.~\eqref{eq:hubbard}, Eq.~\eqref{eq:int-hamiltonian-spin} with $t=0$, and $\hat{H}_\mathrm{m}$ in Eq.~\eqref{eq:charge-hamiltonian}.

First, we discuss Zeeman field modulation perpendicular to the static field, i.e. $\mathbf{\Delta} = (\Delta_x, 0,0)$. 
In this case $[H_\mathrm{spin},H^\mathrm{s}_\mathrm{int}] \neq 0$, 
leading to leakage from the computational state $\ket{\downarrow\downarrow}$ to the state $\ket{\downarrow\uparrow}$, which is outside of the computational subspace (see SM~S4\cite{SuppMat}).
The leakage probability 
for a pre-measurement state $\rho_\mathrm{pre}$ of the computational subspace $\mathrm{span}\left(\left\{\ket{\downarrow\downarrow}, \ket{S_{(0,2)}}\right\}\right)$ at the readout point is expressed as 
\begin{equation}
  \label{eq:leakage-def}
    \mathcal{L}(\rho_\mathrm{pre}) = 
    \mathrm{Tr}\left(P_\mathrm{leak} \mathcal{M}[\rho_\mathrm{pre}]\right),
\end{equation}
where $\mathcal{M} = \mathcal{M}_\downarrow + \mathcal{M}_\uparrow$, and
$P_\mathrm{leak} = 1 - \ket{\downarrow \downarrow}\bra{\downarrow \downarrow} - \ket{S_{(0,2)}}\bra{S_{(0,2)}}$.

Of the two computational basis states, $\ket{S_{(0, 2)}}$ is not prone to readout-induced leakage, but $\ket{\downarrow \downarrow}$ is.
Numerical results for leakage from
$\ket{\downarrow \downarrow}$
are shown with solid lines in Fig.~\ref{fig:spin-setup}f.
Longer integration times lead to increased leakage, saturating at $\mathcal{L}(\tau_\mathrm{int} \to \infty) = 1/2$;
increasing the coupling strength $\Delta_x$ also increases leakage.
Analytical derivations (SM S4 \cite{SuppMat}) for this case show that leakage can be expressed as $\mathcal{L}(\ket{\downarrow\downarrow}) = \left(1-e^{-\Gamma_\mathrm{leak}\tau}\right)/2$, with the leakage rate
\begin{eqnarray}
    \Gamma_\mathrm{leak} &=& \frac{8\Delta_x^2 }{Z^2_R \Delta\tau} \sin^2\left(\frac{Z_R \Delta\tau}{2\hbar}\right).
    \label{eq:leakage-formula}
\end{eqnarray}
This analytical result for the leakage is shown in Fig.~\ref{fig:spin-setup}f as dashed lines, matching well the numerical results.
This leakage mechanism, although undesired as it corrupts the post-measurement state, does not affect the readout infidelity, as the sensor does not differentiate between  $\ket{\downarrow\downarrow}$ and $\ket{\downarrow\uparrow}$.

Second, we discuss Zeeman field modulation parallel to the static field, i.e. $\mathbf{\Delta} = (0,0,\Delta_z)$.
Then, it holds that $[\hat{H}_\mathrm{spin}, \hat{H}_\mathrm{int}^\mathrm{s}] = 0$,
hence, there is no relaxation or leakage, and the infidelity goes to zero as $\tau_\mathrm{int}\rightarrow \infty$. 
This is shown numerically in Fig.~\ref{fig:spin-setup}e,
as the solid lines.
Fig.~\ref{fig:spin-setup}e also reveals that for a given integration time, infidelity decreases and hence readout improves as $\Delta_z$ is increased.
This improvement is due to the spin-charge coupling term of Eq.~\eqref{eq:int-hamiltonian-spin}, which, in this special case, makes the QPC current dependent on  $\hat{s}_{R,z}$ and hence facilitates readout.
This is explained by expressing $\hat{H}_\mathrm{int}^\mathrm{s}$
in the two-dimensional computational subspace, 
resulting in
$-(\delta \gamma+ \Delta_z) \ket{S_{(0,2)}}\bra{S_{(0,2)}}$, up to a global energy shift.
Therefore, we substitute  $\delta \gamma \to \delta \gamma + \Delta_z$ in Eq.~\eqref{eq:measurementrate}, and conclude that $\Gamma_\mathrm{m}$ shows an approximately linear dependence on $\Delta_z$ for $\Delta_z \ll \delta \gamma$.
This is confirmed by the inset of Fig.~\ref{fig:spin-setup}e, where the above analytical approximation (solid) is compared to results (dots) obtained by fitting the function
$1-\Phi\left(\sqrt{2\Gamma_\mathrm{m} \tau_\mathrm{int}}\right)$ on the  infidelity data in Fig.~\ref{fig:spin-setup}e.

The above results imply that leakage is eliminated in a magnetic field configuration where the static Zeeman field in dot $R$ is parallel to the local Zeeman field fluctuation caused the QPC.
Interestingly, this can be achieved for any generic $g$-tensor and $g'$ matrix: the equation $g \mathbf{B} \parallel g' \mathbf{B}$ for the unknown $\mathbf{B}$ is solved by the right eigenvectors of $g^{-1}g'$, and the latter $3\times 3$ matrix has either 1 or 3 real eigenvalue-eigenvector pairs, which describe physical magnetic-field directions \cite{Sen_2023}. 
In the case of 3 real eigenvectors, it is optimal to orient the magnetic field along that of the greatest eigenvalue, to maximize the measurement rate $\Gamma_\mathrm{m}$.
The above consideration identifies a readout sweet spot in terms of the magnetic-field direction, akin to dephasing and relaxation sweet spots identified earlier \cite{Piot, Sen_2023, Michal_2023, Mauro_2024, PhysRevLett.127.190501, PRXQuantum.2.010348, PhysRevLett.129.247701}.
The red arrows in Figs.~\ref{fig:spin-setup}b,c show the readout sweet spot for a the $g$-tensor and $g$-tensor modulation $g'$ from \cite{Crippa_2018}, for which there is a single real eigenvalue of $g^{-1}g'$.
In a multi-qubit device, in-situ tuning of the $g$-tensor parameters by gate voltages \cite{Crippa_2018,Piot,Hendrickx,John,RimbachRuss, PRXQuantum.2.010348, PhysRevLett.129.066801, PhysRevB.104.115425, saezmollejo, carballido, BassiArxiv,Seidler} could be used to synchronize the readout sweet spots 
(see also discussion in SM~S9\cite{SuppMat}). We propose to find the readout sweet spot device configuration by measuring the leakage probablity, as described in SM~S8\cite{SuppMat}.

\emph{Conclusions.}
We have identified two readout error sources for quantum-dot-based charge and spin qubits: residual tunneling and $g$-tensor modulation. 
We have characterized the parameter dependence of these errors via a minimal model of the readout process, going beyond readout infidelity, describing post-measurement state mixedness and leakage - key benchmarks for protocols recycling qubits.
Our results provide clear guidelines to optimize readout by controlling the tunnel coupling and exploiting magnetic readout sweet spots, and pave the way for understanding and minimizing measurement errors across a variety of charge-sensing techniques. 

\emph{Code and data vailability.} Python code used for simulations and plotting, as well as output data obtained from our simulations, are published under Ref.~\onlinecite{svastits_2025_16024845}.

\emph{Acknowledgments.}
We thank Gy.~Frank, B.~Kolok, R.~N\'emeth, Sudipto Das, E.~G.~Kelly, P.~Harvey-Collard for fruitful discussions. 
This work was supported by the Ministry of Culture and Innovation and the National Research, Development and Innovation Office within the Quantum Information National Laboratory of Hungary, Grant No. 2022-2.1.1-NL-2022-00004 (DS, GSz, AP); 
via the EKÖP\_KDP-24-1-BME-2 funding scheme through Project Nr.~2024-2.1.2-EKÖP-KDP-2024-00005 (DS, AP); 
and by the European Union within the Horizon Europe research and innovation programme via the projects IGNITE (DS, AP), ONCHIPS (DS, AP), and QLSI2 (DS, SB, AP).
This work was supported by the HUN-REN Hungarian Research Network through the Supported Research Groups Programme, HUN-REN-BME-BCE Quantum Technology Research Group (TKCS-2024/34).
GSz was supported by the J\'{a}nos Bolyai Research Scholarship of the Hungarian Academy of Science.
BH, SB, and DL also acknowledge support from NCCR Spin (grant number 225153) and the NCCR Spin Mobility Grant.
SB was sponsored in part by the Army Research Office, Award Number: W911NF-23-1-0110. The views and conclusions contained in this document are those of the authors and should not be interpreted as representing the official policies, either expressed or implied, of the Army Research Office or the U.S. Government. The U.S. Government is authorized to reproduce and distribute reprints for Government purposes notwithstanding any copyright notation herein.

\bibliographystyle{apsrev4-2}
\bibliography{ref}

\pagebreak
\clearpage

\onecolumngrid
\vspace*{0.1cm}
\begin{center}
	\large{\bf Supplementary Material to \\`Readout sweet spots for spin qubits with strong spin-orbit interaction'\\}
\end{center}
\begin{center}
	Domonkos Svastits$^{1, 2}$, Bence Hetényi$^3$, Gábor Széchenyi$^4$, James \\ Wotton$^{3, 5}$, Daniel Loss$^6$, Stefano Bosco$^7$, and András Pályi$^{1, 8}$ \\
	{\it\small
    $^1$Budapest University of Technology and Economics, Műegyetem rkp. 3., H-1111 Budapest, Hungary \\
    $^2$Qutility @ Faulhorn Labs, Budapest, Hungary\\
    $^3$IBM Research Europe – Zurich, Säumerstrasse 4, 8803 Rüschlikon, Switzerland \\
    $^4$ELTE Eötvös Loránd University, Institute of Physics, H-1117 Budapest, Hungary \\
    $^5$Moth Quantum AG, Schorenweg 44B, 4144, Arlesheim, Switzerland \\
    $^6$Department of Physics, University of Basel, Klingelbergstrasse 82, CH-4056 Basel, Switzerland\\
    $^7$QuTech and Kavli Institute of Nanoscience, Delft University of Technology, Delft, The Netherlands\\
    $^8$HUN-REN-BME-BCE Quantum Technology Research Group, Műegyetem rkp. 3., H-1111 Budapest, Hungary\\
	(Dated: \today)}
\end{center}
\vspace*{0.1cm}

\twocolumngrid

\setcounter{equation}{0}
\setcounter{figure}{0}
\setcounter{table}{0}
\setcounter{section}{0}

\renewcommand{\theequation}{S\arabic{equation}}
\renewcommand{\thefigure}{S\arabic{figure}}
\renewcommand{\thesection}{S\arabic{section}}
\renewcommand{\bibnumfmt}[1]{[S#1]}

\section{Details of describing QPC-based charge-sensing with the QMQ model}
\label{sec:details-of-QMQ}

We introduced the Qubit Measures Qubit (QMQ) model in the main text. Here, we describe the model in more detail using the framework of indirect measurements introduced in Section 1.2.3 of Ref.~\onlinecite{Wiseman_Milburn_2009}. In this framework, the system is coherently coupled to a meter (or apparatus), and the meter is measured in a projective manner. According to the underlying physical picture of the Landauer-Büttiker formalism, which is the standard description of quantum transport, electrical conduction inside a QPC happens via phase coherent dynamics of electrons. This makes the framework of indirect measurements well suited to describe QPC-based charge sensing.

For QPC-based charge sensing, the system is the qubit hosted by the DQD, the meter is the QPC itself and the coupling is the Coulomb interaction between the electrons inside the DQD and the electrons passing through the QPC. Due to this coupling, the transmission probability of electrons depends on the charge configuration of the DQD. In the QMQ model, the two electronic reservoirs connected by the QPC are replaced by two quantum dots. Movement of electrons between these dots is described by the Hamiltonian $\hat{H}_\mathrm{m}$ (see Eq.~(1)). The dependence of the transmission probability of an electron from one dot to the other is taken into account by the effective coupling term $\hat{H}_\mathrm{int}^\mathrm{c/s}$ (see Eqs.~(1) and (7)).

Our motivation for using the framework of indirect measurements is to describe the time-dependent and stochastic nature of data acquisition and measurement back-action within a single shot, which are central components of experiments.

The remainder of SM~\ref{sec:details-of-QMQ} is structured as follows: In SM~\ref{sec:meas-ops}, we show how to calculate the measurement operators and measurement operations describing a single electron passing the QPC, i.e. describing a single \textit{indirect measurement}. In SM~\ref{sec:weak-m-series}, we calculate the measurement operations corresponding to multiple indirect measurements. In SM~\ref{sec:inference-rule}, we show the inference rule used to infer a single bit (e or g) from the measured current. In SM~\ref{sec:post-m-state}, we give the formulas to calculate the conditional post-measurement states and visualize them on the Bloch-sphere for specific examples. Finally, in SM~\ref{sec:numerical-method}, we describe the efficient numerical method used to calculate conditional dynamics predicted by the QMQ model.

\subsection{Measurement operators of a single indirect measurement}
\label{sec:meas-ops}
Consider charge-qubit readout modeled with the QMQ model, as described in the main text.
Our model describes the readout process as a sequence of indirect measurements.
The conditional post-measurement states of the qubit after a single indirect measurement are
\begin{subequations}
\begin{align}
    \hat{\rho}^\mathrm{post}_0 = \Tr_\mathrm{m}[\ket{B}\bra{B}\hat{U}_\mathrm{tot}(\hat{\rho}_\mathrm{pre}\otimes\ket{B}\bra{B})\hat{U}^\dagger_\mathrm{tot}\ket{B}\bra{B}],\\
    \hat{\rho}^\mathrm{post}_1 = \Tr_\mathrm{m}[\ket{T}\bra{T}\hat{U}_\mathrm{tot}(\hat{\rho}_\mathrm{pre}\otimes\ket{B}\bra{B})\hat{U}^\dagger_\mathrm{tot}\ket{T}\bra{T}],
\end{align}
\label{eq:rho-post}
\end{subequations}
given that the measurement of the meter resulted in a projection into the meter-state $\ket{B}$ and $\ket{T}$ respectively. Here, $\hat{U}_\mathrm{tot}=e^{-\frac{i}{\hbar}\hat{H}_\mathrm{tot}\Delta \tau}$ denotes unitary time evolution according to the Hamiltonian $\hat{H}_\mathrm{tot}$ in Eq.~(1) for the charge qubit; and the corresponding total Hamiltonian for the spin qubit. Furthermore, $\Tr_\mathrm{m}[...]$ denotes tracing out the meter. 

Simplifying Eq.~\eqref{eq:rho-post}, we obtain
\begin{equation}
    \hat{\rho}^\mathrm{post}_{0/1} \equiv \mathcal{M}_{0/1}[\hat{\rho}_\mathrm{pre}] =\hat{M}_{0/1}\hat{\rho}_\mathrm{pre}\hat{M}_{0/1}^\dagger.
\end{equation}
Here, we defined the measurement operations
$\mathcal{M}_{0}$ and $\mathcal{M}_{1}$ as well as the measurement operators \cite{Wiseman_Milburn_2009}
\begin{subequations}
\begin{align}
    &\bra{i}\hat{M}_{0}\ket{j} = \bra{i, B} \hat{U}_\mathrm{tot} \ket{j, B}, \\
    &\bra{i}\hat{M}_{1}\ket{j} = \bra{i, T} \hat{U}_\mathrm{tot} \ket{j, B}, 
    \label{eq:meas-ops-def}
\end{align}
\end{subequations}
where the index $i, j \in \{e, g\}$ labels the states of the qubit.
The measurement operator $\hat{M}_0$ ($\hat{M}_1$) corresponds to finding the QPC-electron in the bottom (top) quantum dot.

To illustrate the measurement operators, we calculate their matrices for the charge qubit with $t=0$ in the computational basis $\{\ket{\mathrm{e}}, \ket{\mathrm{g}}\}$:
\begin{subequations}
\begin{align}
    M_r &= U\Tilde{M}_r, \text{ with}\\
    U &= \begin{pmatrix}
        \exp\left(-\frac{i\epsilon \Delta\tau}{\hbar}\right) & 0 \\
        0 & \exp\left(\frac{i\epsilon \Delta\tau}{\hbar}\right)
    \end{pmatrix},\\
    \Tilde{M}_0 &= \begin{pmatrix}
       \sqrt{1-p_{\mathrm{e}}} & 0 \\
        0 & \sqrt{1-p_{\mathrm{g}}}
    \end{pmatrix},\\
    \Tilde{M}_1 &= \begin{pmatrix}
       \sqrt{p_{\mathrm{e}}} & 0 \\
        0 & \sqrt{p_{\mathrm{g}}}
    \end{pmatrix},\\
    p_{\mathrm{e}}&=\sin^2\left(\frac{\gamma\Delta\tau}{\hbar}\right), \\
    p_{\mathrm{g}} &= \sin^2\left(\frac{(\gamma-\delta\gamma)\Delta\tau}{\hbar}\right).
\end{align}
\label{eq:meas-ops0}
\end{subequations}

In this work, we tune the parameters of the QMQ model in such a way that the probability $p_{\mathrm{e}}$ ($p_{\mathrm{g}}$) of transmission from the bottom dot to the top dot is slightly more (less) than 0.5, given that the data qubit electron is in the left (right) dot. 
Therefore, in the $t=0$ case above, a consequence of the formulas is that as follows: if a single indirect measurement  it is easy to understand the effect of a single indirect measurement. For example, if we measure 0, the amplitude of being in the left dot decreases relative to the amplitude of being in the right dot. Recall that for $t=0$, we have $\ket{L} = \ket{\mathrm{e}}$ and $\ket{R} = \ket{\mathrm{g}}$.

It is often useful in both analytical (see SM \ref{sec:dephasing-meas-rates} and \ref{sec:rel-rate}) and numerical (see SM~\ref{sec:numerical-method}) calculations to use the transfer matrix representation $\Upsilon_{0/1}$ of the measurement operation $\mathcal{M}_{0/1}$. The transfer matrix $\Upsilon_{0/1}$ acts on the vectorized form of the density matrix $(\rho_{ee}, \rho_{ge}, \rho_{eg}, \rho_{gg})^T$ \cite{hashim2024practicalintroductionbenchmarkingcharacterization} and it can be expressed using the matrix representation of the measurement operators as
\begin{equation}
    \Upsilon_{0/1} = M_{0/1}^* \otimes M_{0/1}.
    \label{eq:transfer-m}
\end{equation}
Equivalently, the transfer matrix elements can be expressed as $(\Upsilon_{0/1})_{\overline{ik},\overline{jl}} = M_{ij}^* M_{kl}$. Here we introduced the notation $\overline{ik} \equiv i\cdot d + k$,  where $d$ is the dimension of the Hilbert space of the data qubit system. For the charge qubit $d=2$, therefore the possible indices are 0, 1, 2 and 3. Note that the transfer matrices depend on the basis in which the measurement operators and density matrices are written. Throughout this work, we use the energy eigenbasis $\{\ket{\mathrm{e}}, \ket{\mathrm{g}}\}$.

\subsection{Series of indirect measurements}
\label{sec:weak-m-series}
As long as the interaction of the qubit and the meter is weak, many indirect measurements are needed to be able to distinguish the computational states $\ket{\mathrm{e}}$ and $\ket{\mathrm{g}}$.
The outcome of a series of $N$ indirect measurements is a bitstring $\mathbf{r}$ of length $N$. The corresponding measurement operator is
\begin{equation}
    \hat{M}_{\mathbf{r}} \equiv \hat{M}_{r_N} ... \hat{M}_{r_2} \hat{M}_{r_1}.
    \label{eq:Mr}
\end{equation}
To provide a full description of the measurement, one needs to evaluate the measurement operators corresponding to all $2^N$ possible bitstrings.
The calculation and storage of the corresponding $2^N$ measurement operators is inefficient.

Although calculating all the measurement operators is in practice impossible for large $N$, interesting quantities can still be calculated. In experiments, the initial state of the qubit is usually inferred based on the average current \cite{PhysRevLett.134.023601, PhysRevX.8.021046, PRXQuantum.3.010352, PRXQuantum.4.010329}, i.e. the total number of transmitted electrons $N_\mathrm{t}$, which contains less information then the time dependence of the current, which is represented by $\mathbf{r}$. For our purposes, it is more convenient to use the transmission ratio $k\equiv N_\mathrm{t}/N$ instead of $N_\mathrm{t}$ as the outcome of the measurement. The number of transmitted electrons $N_\mathrm{t}$ and the transmission ratio both can take $N+1$ different values. The corresponding measurement operations are
\begin{equation}
    \mathcal{M}_k^{(N)}[\hat{\rho}] = \sum_{\frac{1}{N}\sum_{i=1}^N r_i = k}\hat{M}_{\mathbf{r}} \hat{\rho} \hat{M}_{\mathbf{r}}^\dagger.
    \label{eq:M_K}
\end{equation}
Note, that if $\hat{M}_0$ and $\hat{M}_1$ commute, the map $\mathcal{M}_k^{(N)}$ takes pure states to pure states, i.e. the measurement is efficient \cite{Wiseman_Milburn_2009}. In this case, no information is lost if only the transmission ratio is stored instead of the bitstring $\mathbf{r}$. We show in SM~\ref{sec:numerical-method} how the measurement operations $\mathcal{M}_k^{(N)}$ are efficiently computable numerically.

\begin{figure*}
    \centering
    \includegraphics[width=\linewidth]{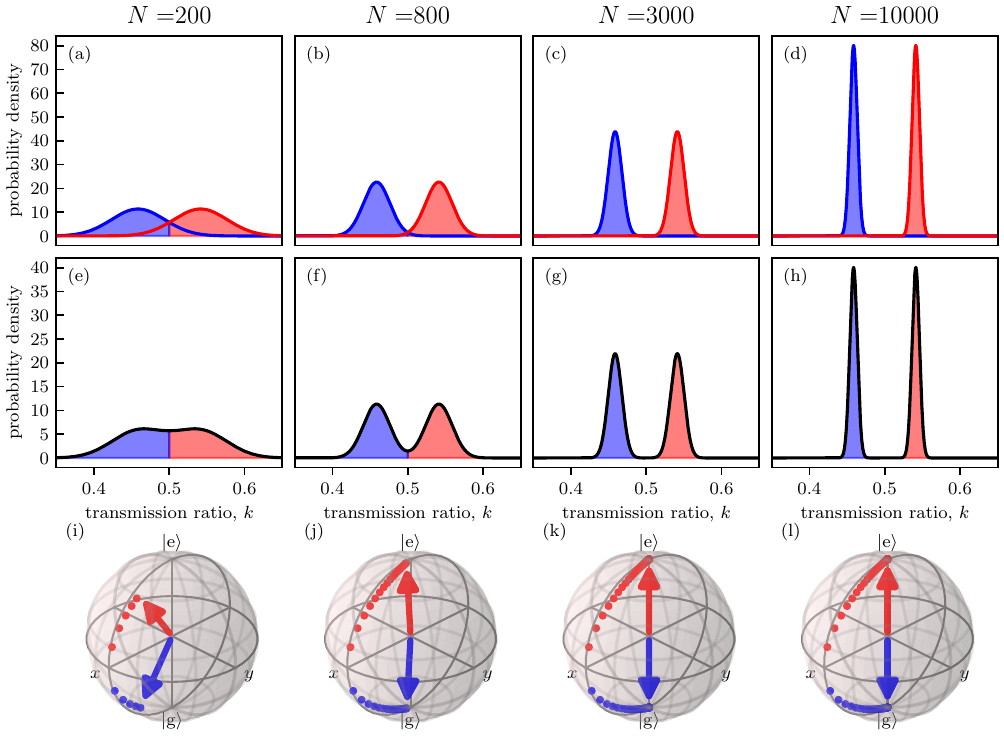}
    \caption{Current distributions and post-measurement states with zero residual tunneling. (a)-(d) Distributions of the average current for the initial states $\ket{\mathrm{e}}$ (red) and $\ket{\mathrm{g}}$ (blue). The red (blue) coloring of the areas under the curves indicates a final outcome of e (g). For example, the part of blue area that is under the red curve is equal to the inference error $\epsilon_{\downarrow}$. (e)-(h) Distribution of the average current for the initial state $\ket{+}=(\ket{\mathrm{e}}+\ket{\mathrm{g}})/\sqrt{2}$. (i)-(l) Post-measurement states for the initial state $\ket{+}$ in the rotating frame of the charge qubit Hamiltonian $\hat{H}_\mathrm{charge}$. The red and blue arrows represent the post-measurement states $\hat{U}^\dagger\mathcal{M}_{\mathrm{e}}[\ket{+}\bra{+}]\hat{U}$ and $\hat{U}^\dagger\mathcal{M}_{\mathrm{g}}[\ket{+}\bra{+}]\hat{U}$ respectively with $\hat{U}= e^{-i\hat{H}_\mathrm{charge}N\Delta \tau}$. The red and blue dots show the post-measurement states for $N=40\cdot i$, $i\in\mathbb{N}$. The parameters used for the simulations are $\epsilon=10$~$\mathrm{\mu eV}$, $t=0$~$\mathrm{\mu eV}$, $\gamma=5$~$\mathrm{\mu eV}$, $\delta\gamma = 0.5$~$\mathrm{\mu eV}$.}
    \label{fig:bloch-spheres0}
\end{figure*}

\begin{figure*}
    \centering
    \includegraphics[width=\linewidth]{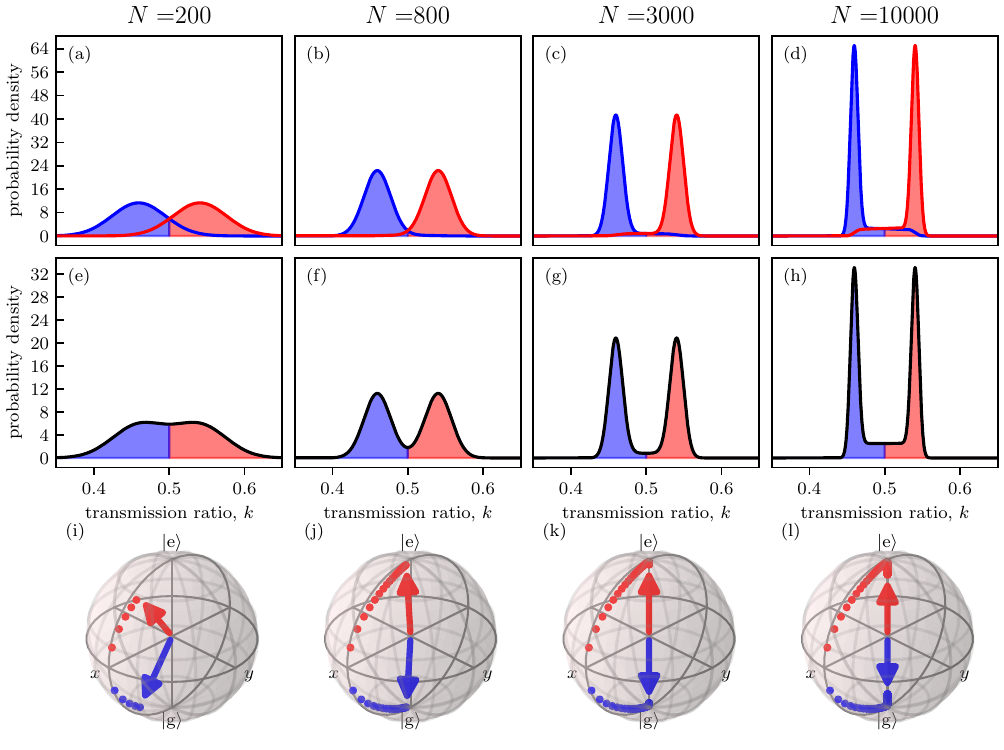}
    \caption{Current distributions and post-measurement states with finite residual tunneling. (a)-(d) Distributions of the average current for the initial states $\ket{\mathrm{e}}$ (red) and $\ket{\mathrm{g}}$ (blue). The red (blue) coloring of the areas under the curves indicates a final outcome of e (g). For example, the part of blue area that is under the red curve is equal to the inference error $\epsilon_{\downarrow}$. (e)-(h) Distribution of the average current for the initial state $\ket{+}=(\ket{\mathrm{e}}+\ket{\mathrm{g}})/\sqrt{2}$. (i)-(l) Post-measurement states for the initial state $\ket{+}$ in the rotating frame of the charge qubit Hamiltonian $\hat{H}_\mathrm{charge}$. The red and blue arrows represent the post-measurement states $\hat{U}^\dagger\mathcal{M}_{\mathrm{e}}[\ket{+}\bra{+}]\hat{U}$ and $\hat{U}^\dagger\mathcal{M}_{\mathrm{g}}[\ket{+}\bra{+}]\hat{U}$ respectively with $\hat{U}= e^{-i\hat{H}_\mathrm{charge}N\Delta \tau}$. The red and blue dots show the post-measurement states for $N=40\cdot i$, $i\in\mathbb{N}$. The parameters used for the simulations are $\epsilon=10$~$\mathrm{\mu eV}$, $t=2$~$\mathrm{\mu eV}$, $\gamma=5$~$\mathrm{\mu eV}$, $\delta\gamma = 0.5$~$\mathrm{\mu eV}$.}
    \label{fig:bloch-spheres}
\end{figure*}

\subsection{Inference rule}
\label{sec:inference-rule}
The outcome of the readout is a time dependent current, which is represented as a bitstring $\mathbf{r}$ in our model. Our aim is to infer the initial state based on the time dependent current. Therefore, we need an inference rule that assigns a single bit to $\mathbf{r}$. A possible approach would be to use the maximum likelihood inference rule based on all the available information in the model. This would require calculating the probabilities $P_{\mathrm{e}}(\mathbf{r})$ and $P_{\mathrm{g}}(\mathbf{r})$ of measuring $\mathbf{r}$ given that the initial state was $\ket{\mathrm{e}}$ and $\ket{\mathrm{g}}$ respectively. Then, we would infer that the initial state was $\ket{\mathrm{e}}$ if $P_{\mathrm{e}}(\mathbf{r}) > P_{\mathrm{g}}(\mathbf{r})$; otherwise we would infer $\ket{\mathrm{g}}$. However, calculating $P_{\mathrm{e}}(\mathbf{r})$ and $P_{\mathrm{g}}(\mathbf{r})$ for all $2^N$ possible different bitstrings $\mathbf{r}$ is inefficient.

Instead, we use the maximum likelihood inference rule, based on only the transmission ratio $k$. This is motivated by experiments, where the average current is used to make this inference. To determine this inference rule, we need to calculate which initial state results in a higher probability of obtaining the transmission ratio $k$, i.e., whether $P_{\mathrm{e}}(k)$ or $P_{\mathrm{g}}(k)$ is bigger. These probabilities are straightforward to calculate:
\begin{subequations}
\begin{align}
    P_{\mathrm{e}}(k) = \Tr[\mathcal{M}_k^{(N)}[\ket{\mathrm{e}}\bra{\mathrm{e}}]],\\
    P_{\mathrm{g}}(k) = \Tr[\mathcal{M}_k^{(N)}[\ket{\mathrm{g}}\bra{\mathrm{g}}]].
\end{align}
\label{eq:P_k}
\end{subequations}
The resulting inference rule is the following: there is a critical transmission ratio $k_c$, above (below) which we infer e (g). The probability mass functions $P_{\mathrm{e}}(k)$ and $P_{\mathrm{g}}(k)$ are efficiently calculable, therefore, so is the critical transmission ratio $k_c$. 
In Figs.~\ref{fig:bloch-spheres0}a-d and \ref{fig:bloch-spheres}a-d, the probabilities $P_{\mathrm{e}}(k)$ and $P_{\mathrm{g}}(k)$ are plotted for different $N$ to illustrate the inference rule. The areas under the curves are colored red (blue) above (below) $k_c$.
Fig.~\ref{fig:bloch-spheres0} corresponds to the case of zero tunneling in the charge qubit ($t=0$), and Fig.~\ref{fig:bloch-spheres} corresponds to nonzero tunneling ($t >0$).

In the $t=0$ case, the critical transmission ratio is easily determined analytically. Due to fixing $\Delta\tau$ as in Eq.~\eqref{eq:timestep0}, we have $p_{\mathrm{e}} = 0.5+\Delta p/2$ and $p_{\mathrm{g}}=0.5-\Delta p/2$; and $p_{\mathrm{e}}$ and $p_{\mathrm{g}}$ are defined in Eq.~\eqref{eq:meas-ops0}. Then, by substituting Eqs.~\eqref{eq:meas-ops0} and \eqref{eq:M_K} into Eq.~\eqref{eq:P_k}, we obtain $k_c = 0.5$.

Furthermore, observe that in the $t=0$ case, $[\hat{M}_0, \hat{M}_1] = 0$ (see Eq.~\eqref{eq:meas-ops0}), meaning that we discard no information by making the inference based on only the transmission ratio, as we have already discussed briefly below Eq.~\eqref{eq:M_K}. If $t\neq 0$, the timing of the transmission events does contain additional information, which is thrown away by this inference rule. It is an interesting open question how this additional information can be used to improve the inference of the initial data qubit state.

\subsection{Post-measurement state}
\label{sec:post-m-state}
Here we show how the post-measurement states corresponding to the inferred value are calculated. As detailed in the previous section, we use maximum likelihood inference based on the transmission ratio $k$ (see SM~\ref{sec:inference-rule}). This results in the following simple rule: we infer e (g) if $k>k_c$ ($k\leq k_c$). Therefore, the measurement operations belonging to the final outcome e and g are
\begin{subequations}
\begin{align}
    \mathcal{M}_{\mathrm{g}}^{(N)} [\hat{\rho}] &= \sum_{k \leq k_c} \mathcal{M}_k^{(N)}[\hat{\rho}]\\
    \mathcal{M}_{\mathrm{e}}^{(N)} [\hat{\rho}] &= \sum_{k > k_c} \mathcal{M}_k^{(N)}[\hat{\rho}].    
\end{align}
\label{eq:meas-operations-eg}
\end{subequations}
These measurement operations give the post-measurement state given that we inferred g or e for the initial state. They can be calculated from the measurement operations $\mathcal{M}_k^{(N)}$ using $\mathcal{O}(N)$ additions.

We plot the transmission ratio distributions and the post-measurement states for the initial state $\ket{+}=(\ket{\mathrm{e}}+\ket{\mathrm{g}})/\sqrt{2}$ in Fig.~\ref{fig:bloch-spheres0}e-l for zero residual tunneling and in \ref{fig:bloch-spheres}e-l for finite residual tunneling. The post-measurement states move towards the two corresponding computational basis states at a time scale $1/\Gamma_\mathrm{m}$. 
For nonzero residual tunneling, relaxation of the qubit state also happens.
This is apparent in Fig.~\ref{fig:bloch-spheres}i-l; for the parameter set used there, the relaxation time scale $1/\Gamma_\mathrm{rel}$ is much longer than $1/\Gamma_\mathrm{m}$.

\subsection{Efficient numerical simulation}
\label{sec:numerical-method}
Here, we show how the measurement operations $\mathcal{M}_k^{(N)}$ in Eq.~\eqref{eq:M_K} are computed efficiently. A brute-force approach would require calculating all $2^N$ measurement operators $M_{\mathbf{r}}$ using Eq.~\eqref{eq:Mr},
and performing the summation in Eq.~\eqref{eq:M_K}. The computational costs of this approach scale exponentially with $N$, making it impossible to carry out for large $N$.

\begin{figure}
    \centering
    \includegraphics[width=\linewidth]{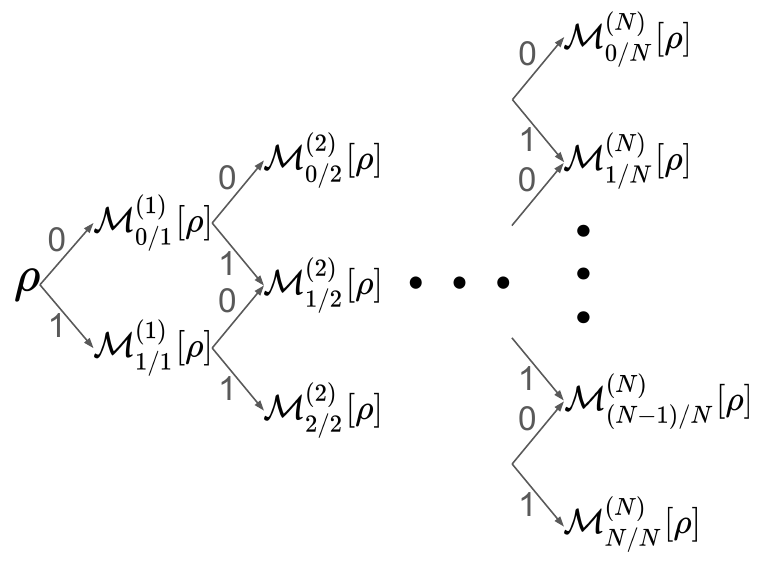}
    \caption{Efficient numerical method, to calculate the conditional time evolution of the data qubit. Superscripts indicate the number of weak measurements, subscripts indicate the transmission ratio $N_\mathrm{t}/N$. The figure shows how the measurement operations of the $N$th round are calculated from the measurement operations of the $(N-1)$th round.}
    \label{fig:tree}
\end{figure}

We instead calculate the measurement operations $\mathcal{M}_k^{(N)}$ directly, using iteration in the number of indirect measurements. 
The iteration step is
\begin{equation}
    \mathcal{M}_{\frac{N_\mathrm{t}}{N}}^{(N)}[\hat{\rho}] = \begin{cases}
    \begin{split}
        &\hat{M}_0\mathcal{M}_\frac{N_\mathrm{t}}{N-1}^{(N-1)}[\hat{\rho}]\hat{M}_0^\dagger +\\ & \  \hat{M}_1\mathcal{M}_\frac{N_\mathrm{t}-1}{N-1}^{(N-1)}[\hat{\rho}]\hat{M}_1^\dagger,\end{split} &\text{for } 1\leq N_\mathrm{t} \leq N-1,\\
        \hat{M}_0\mathcal{M}_\frac{0}{N-1}^{(N-1)}[\hat{\rho}]\hat{M}_0^\dagger, &\text{for } N_\mathrm{t} = 0,\\
        \hat{M}_1\mathcal{M}_\frac{N-1}{N-1}^{(N-1)}[\hat\rho]\hat{M}_1^\dagger, &\text{for } N_\mathrm{t}=N.
    \end{cases}
\end{equation}
The iteration is illustrated in Fig.~\ref{fig:tree}. In the $N$th step of the iteration, $N_\mathrm{t}$ can take $N+1$ different values, so we need to compute $N+1$ measurement operations. Let us call the computation of one measurement operation an elementary computation. Then the total number of elementary computations up to $N$ weak measurements is
\begin{equation}
    \sum_{i=1}^{N} (i+1) = \frac{N(N+3)}{2} = O(N^2).
\end{equation}
Note that the elementary computations of a single round can be parallelized.

In our numerical calculations, we represented the measurement operations as transfer matrices $\Upsilon_k^{(N)}$ and performed the iteration in terms of these.

\section{Relating model parameters to experimental parameters}
\label{sec:exp-params}

In our numerical calculations, we use experimentally constrained parameter values for calculating the measurement operations $\mathcal{M}_{\mathrm{e}}$, $\mathcal{M}_{\mathrm{g}}$; and measurement benchmarks. To do this, we need to match the experimental parameters and the parameters of our model.\par
We fix the timestep between meter measurements as
\begin{equation}
    \Delta \tau = \frac{\hbar \pi}{4\left(\gamma - \frac{\delta\gamma}{2}\right)},
    \label{eq:timestep0}
\end{equation}
which corresponds to 0.5 transmission probability if the data qubit state is an equal superposition or mixture of the states $\ket{L}$ and $\ket{R}$. This choice is motivated by experiments, since the conductance of a QPC  is most sensitive to changes of charge configuration at the point, where its conductance is $G_0/2$, i.e. half the conductance quantum \cite{Vigneau_2023}. In the QPC, there is only one open conductance channel, thus a conductance of  $G_0/2$ means that any time an electron approaches the QPC, it tunnels through with probability $0.5$. For the spin qubit, in the presence of a finite $\Delta_z$ term, the same consideration leads to 
\begin{equation}
    \Delta \tau = \frac{\hbar \pi}{4\left(\gamma - \frac{\delta\gamma-\Delta_z}{2}\right)}.
    \label{eq:timestep}
\end{equation}
With this choice, the transmission probability is 0.5 if the qubit state is the equal superposition or mixture of the two computational states $\ket{S_{(0,2)}}$ and $\ket{\downarrow\downarrow}$.

Next, we calculate the current flowing through the QPC predicted by our model. The average current is
\begin{equation}
    \Bar{I} = \frac{e}{2\Delta\tau},
    \label{eq:avg-current}
\end{equation}
where $e$ is the elementary charge. The difference between the currents corresponding to the states $\ket{L}$ and $\ket{R}$ is 
\begin{equation}
    \Delta \bar{I} = \frac{e}{\Delta\tau} \left[\sin^2\left(\frac{\Delta\tau}{\hbar}\gamma\right)-\sin^2\left(\frac{\Delta\tau}{\hbar}\left(\gamma-\delta\gamma\right)\right)\right]\approx \frac{\delta\gamma \, e}{\hbar}.
    \label{eq:diff-current}
\end{equation}
At the second equation mark, we expanded the sine functions up to first order around $\pi/4$, using Eq.~\eqref{eq:timestep} and the relation $\delta\gamma \ll \gamma$. Eqs.~\eqref{eq:avg-current} and \eqref{eq:diff-current} connect the experimentally measurable currents and the model parameters.\par
The above relations lead us to the following experimentally realistic  parameter set for the charge qubit \cite{PhysRevLett.134.023601}:
\begin{subequations}
\begin{align}
    \gamma  &= 5 \ \mathrm{\mu eV} \Leftrightarrow  \Bar{I}\sim 1 \text{ nA} \\
    \delta\gamma  &= 0.5 \ \mathrm{\mu eV} \Leftrightarrow \Delta \bar{I} \sim 100 \text{pA}\\
    \epsilon  &= 10 \ \mathrm{\mu eV}\\
    t &= 0 - 2 \ \mathrm{\mu eV}.
\end{align}
\end{subequations}
For the spin qubit, we use the same values for the parameters $\gamma$ and $\delta\gamma$ of the QPC. For parameters of the spin-Hamiltonian in Eq.~(6), we used the parameter set
\begin{subequations}
\begin{align}
    U  &= 1 \ \text{meV},\\
    \epsilon  &= U + 40 \ \mathrm{\mu eV}, \\
    Z_L  &= 11 \ \mathrm{\mu eV},\\
    Z_R &= 9 \ \mathrm{\mu eV},\\
    t &= 0,
\end{align}
\end{subequations}
which is experimentally realistic for electrons in AlGaAs/GaAs and Si/SiGe heterostructures \cite{PRXQuantum.4.010329, PRXQuantum.3.010352, PhysRevX.8.021046}.

The integration time in experiments is around $\tau_\mathrm{int}\sim 10-100 \text{\ \textmu s}$. In our model, current sensing is assumed to be perfect, consequently, according to our model, a measurement time much shorter than the above $\tau_\mathrm{int}$ is sufficient to distinguish between the computational basis states of the data qubit. It is possible to incorporate imperfect current sensing into the model, by using inefficient \cite{Wiseman_Milburn_2009} meter measurements, however, we leave this generalization for later work.

\section{Dephasing and measurement rates}
\label{sec:dephasing-meas-rates}
Here, we determine the dephasing and measurement rates for the case of zero tunneling amplitude. For finite, but small tunneling amplitude $t\ll \epsilon$, the derived rates are the zeroth order approximations in $t/\epsilon$ of the actual rates. Here, we do the derivation for the charge qubit, but an analogous derivation works for the spin qubit as well.

The Hamiltonian of the charge qubit with zero tunneling amplitude is $\hat{H}_\mathrm{charge} = \epsilon\hat{\sigma}_z$, which commutes with $\hat{H}_\mathrm{int}^\mathrm{c}$. Consequently, the measurement operators also commute:
\begin{equation}
    [\hat{M}_0, \hat{M}_1] = 0.
    \label{eq:meas_ops_commutation}
\end{equation}
The matrix of these measurement operators in the computational basis is written in Eq.~\eqref{eq:meas-ops0}, from which we get the transfer matrices using Eq.~\eqref{eq:transfer-m}:
\begin{subequations}
\begin{align}
    \Upsilon_0 &= \text{diag}\begin{pmatrix}
    1-p_{\mathrm{e}}\\
    \sqrt{(1-p_{\mathrm{e}})(1-p_{\mathrm{g}})}e^{\frac{i}{\hbar}2\epsilon\Delta\tau}\\
    \sqrt{(1-p_{\mathrm{e}})(1-p_{\mathrm{g}})}e^{-\frac{i}{\hbar}2\epsilon\Delta\tau}\\
    1-p_{\mathrm{g}}\end{pmatrix},\\
    \Upsilon_1 &= \text{diag}\begin{pmatrix}
    p_{\mathrm{e}}\\
    \sqrt{p_{\mathrm{e}} p_{\mathrm{g}}}e^{\frac{i}{\hbar}2\epsilon\Delta\tau}\\
    \sqrt{p_{\mathrm{e}} p_{\mathrm{g}}}e^{-\frac{i}{\hbar}2\epsilon\Delta\tau}\\
    p_{\mathrm{g}}\end{pmatrix},
\end{align}
\label{eq:transfer-m0}
\end{subequations}
Due to Eq.~\eqref{eq:meas_ops_commutation}, we have $[\Upsilon_0, \Upsilon_1]=0$, which is also apparent from Eq.~\eqref{eq:transfer-m0}. Therefore, the transfer matrix representing $\mathcal{M}_k^{(N)}$ is
\begin{equation}
    \Upsilon_k^{(N)} = {N \choose kN} \Upsilon_1^{kN} \Upsilon_0^{(N-kN)}.
\end{equation}
Remember that $kN = N_\mathrm{t}$. This can than be written in the form 
\begin{equation}
    (\Upsilon_k^{(N)})_{ii} = ((\Upsilon_0)_{ii}+(\Upsilon_1)_{ii})^N P_\mathrm{B}\left(kN, N, \frac{(\Upsilon_1)_{ii}}{(\Upsilon_0)_{ii}+(\Upsilon_1)_{ii}}\right),
    \label{eq:transfer-m-ii}
\end{equation}
where $P_\mathrm{B}(kN, N, p)$ is the probability mass function of the binomial distribution. Note that the expression in Eq.~\eqref{eq:transfer-m-ii} makes sense, since the argument $(\Upsilon_1)_{ii} / ((\Upsilon_0)_{ii}+(\Upsilon_1)_{ii})$ of the mass function is a real number between 0 and 1.

The binomial distribution can be approximated by the a normal distribution for $N\gg 1$. Using this approximation and Eq.~\eqref{eq:transfer-m0}, we get
\begin{subequations}
\begin{align}
    \Upsilon_{\mathrm{g}}^{(N)} &\equiv \int_{-\infty}^{k_c} \Upsilon_k^{(N)}dk = \text{diag}(\epsilon_\downarrow, \alpha, \alpha^*, 1-\epsilon_\uparrow),\\
    \Upsilon_{\mathrm{e}}^{(N)} &\equiv \int_{k_c}^{\infty} \Upsilon_k^{(N)}dk = \text{diag}(1-\epsilon_\downarrow, \beta, \beta^*, \epsilon_\uparrow),
\end{align}
\label{eq:transfer-m-qnd1}
\end{subequations}
where
\begin{subequations}
\begin{align}
    \label{eq:transfer-m-eps}
    \epsilon_\uparrow(N) &= \epsilon_\downarrow(N)  = \Phi\left(\frac{-\Delta p}{\sqrt{1-\Delta p^2}}\sqrt{N}\right),\\
    \label{eq:transfer-m-alpha}
    \alpha(N) &= \beta(N) = \frac{1}{2}\sqrt{1-\Delta p^2}^N e^{\frac{i}{\hbar}2\epsilon N\Delta\tau}.
\end{align}
\end{subequations}
Here, $\Phi$ denotes the cumulative distribution function of the normal distribution. In Eq.~\eqref{eq:transfer-m-qnd1}, we omitted the arguments $N$ for brevity.
The matrix element $\epsilon_\uparrow$ ($\epsilon_\downarrow$) gives the probability of inferring g (e) for the initial state $\ket{\mathrm{e}}$ ($\ket{\mathrm{g}}$); $\alpha$ and $\beta$ give how the off-diagonal density matrix elements evolve.

We now define the measurement and dephasing rates using Eqs.~\eqref{eq:transfer-m-eps} and \eqref{eq:transfer-m-alpha} respectively. The measurement rate characterizes the speed at which the two current distributions corresponding to the initial states $\ket{\mathrm{e}}$ and $\ket{\mathrm{g}}$ become distinguishable. We define it through
\begin{equation}
    \epsilon_\uparrow = \epsilon_\downarrow \equiv \Phi\left(-\sqrt{2\Gamma_\mathrm{m} \tau_\mathrm{int}}\right).
    \label{eq:meas-rate-def}
\end{equation}
The value of the measurement rate is obtained by comparing this definition with Eq.~\eqref{eq:transfer-m-eps}:
\begin{equation}
    \Gamma_\mathrm{m} = \frac{1}{\Delta \tau}\frac{\Delta p^2}{2(1-\Delta p^2)} \approx \frac{1}{2} \left(\frac{\delta\gamma}{\hbar}\right)^2 \Delta \tau. \label{eq:meas-rate-val}
\end{equation}
At the second equation mark in Eq.~\eqref{eq:meas-rate-val} we expanded $\Gamma_\mathrm{m}$ up to second order in $\delta\gamma/\gamma$ at the readout point given by Eq.~\eqref{eq:timestep0}. For the spin qubit, the same calculation yields
\begin{subequations}
\begin{align}
    \Gamma_\mathrm{m} &= \frac{1}{\Delta \tau}\frac{\Delta p_\mathrm{s}^2}{2(1-\Delta p_\mathrm{s}^2)} \approx \frac{1}{2} \left(\frac{\delta\gamma+\Delta_z}{\hbar}\right)^2 \Delta \tau, \label{eq:meas-rate-val-spin}\\
    \Delta p_\mathrm{s} &= \sin^2\left(\frac{(\gamma + \Delta_z)\Delta\tau}{\hbar}\right) - \sin^2\left(\frac{(\gamma -\delta\gamma)\Delta\tau}{\hbar}\right),
\end{align}
\end{subequations}
where $\Delta p_\mathrm{s}$ is the analogous quantity to $\Delta p$ for the spin qubit, i.e it is the difference of the transition probabilities for the states $\ket{T_-}$ and $\ket{S_{(0, 2)}}$.
For $\Delta_z = 0$ this is exactly the same as the measurement rate for the charge qubit Eq.~\eqref{eq:meas-rate-val}.

The dephasing $\Gamma_\mathrm{d}$ rate characterizes the speed at which the magnitude of the off-diagonal matrix elements decrease with time in the unconditional time evolution, i.e. it is defined through
\begin{equation}
    \abs{\rho_{eg}}\sim e^{-\Gamma_\mathrm{d} \tau_\mathrm{int}}. \label{eq:dep-rate-def}
\end{equation}
Its value is obtained by comparing this definition with Eq.~\eqref{eq:transfer-m-alpha}:
\begin{equation}
\begin{split}
    \Gamma_\mathrm{d} &= -\frac{\ln\left(\abs{\alpha(1)+\beta(1)}\right)}{\Delta\tau} \\
    &= -\frac{1}{2\Delta\tau}\ln\left(1-\Delta p^2\right) \approx \frac{1}{2}\left(\frac{\delta\gamma}{\hbar}\right)^2 \Delta\tau \label{eq:dep-rate-val}.
\end{split}
\end{equation}
Similarly to the measurement rate, we expanded the dephasing rate up to second order in $\delta\gamma/\gamma$. We find, that up to this order, the two rates are the same for the charge qubit. For the spin qubit, an analogous calculation shows that for $\Delta_x = 0$, the off-diagonal density matrix elements between the two computational states decay as
\begin{equation}
    \abs{\rho_{S_{(0,2), \downarrow\downarrow}}} \sim e^{-\Gamma_\mathrm{d} \tau_\mathrm{int}},
\end{equation}
with 
\begin{equation}
    \Gamma_\mathrm{d} = -\frac{1}{2\Delta\tau}\ln\left(1-\Delta p_s^2\right) \approx \frac{1}{2}\left(\frac{\delta\gamma+\Delta_z}{\hbar}\right)^2 \Delta\tau.
\end{equation}
The time evolution of off-diagonal matrix elements connecting states outside of the computational subspace can be calculated analogously.

Our results show that the decrease of the off-diagonal density matrix elements and the separation of the two current distributions happen at the same time scale. Note that this approximate equality of the two rates holds only because we assume current sensing to be perfect in our model. Noisy current sensing does not change the dephasing rate, since it is defined through the unconditional time evolution. However, it decreases the measurement rate, i.e., increases the time needed to distinguish between the two computational states. Therefore, in practice, the dephasing rate sets an upper bound for the measurement rate.

\section{Relaxation rate}
\label{sec:rel-rate}
Here, we derive the relaxation rate Eq.~(4) for the charge qubit. An analogous derivation results in the leakage rate Eq.~(10) for the spin qubit. In the derivation, we use perturbation theory to determine the transfer matrix $\Upsilon$ describing the unconditional time evolution of one timestep. In the charge qubit case, the transfer matrix is a $4\times 4$ matrix, which has an eigenvalue that is 1. Its second closest eigenvalue to 1 in terms of absolute value determines the speed at which the initial state approaches the steady state, therefore, it is related to the relaxation rate as $e^{-\Gamma_\mathrm{rel}^\mathrm{c}\Delta\tau} = \abs{\lambda_\mathrm{rel}}$. In the calculation, we determine the eigenvalue $\lambda_\mathrm{rel}$ and from it the relaxation rate $\Gamma_\mathrm{rel}^\mathrm{c}$ up to leading order in $t/\epsilon$, $\delta\gamma/\epsilon$ and $\delta\gamma/\gamma$.

The derivation uses first order time dependent perturbation theory. The unperturbed Hamiltonian is $\hat{H}_0 = \hat{H}_\mathrm{charge} + \hat{H}_\mathrm{m}$, while the perturbation is the interaction Hamiltonian $\hat{H}_\mathrm{int}^\mathrm{c}$ (see Eq.~(1)). The first step is to determine the eigenvectors of $\hat{H}_0$ and rewrite the interaction Hamiltonian $\hat{H}_\mathrm{int}^{\mathrm{c}}$ in the eigenbasis of $\hat{H}_0$. Eigenvectors of $\hat{H}_0$ can be written in the form $\ket{e_i,s}\equiv \ket{e_i}\otimes\ket{s}$, where $\ket{e_i}\in \{\ket{\mathrm{e}}, \ket{\mathrm{g}}\}$ is an eigenvector of $\hat{H}_\mathrm{charge}$ and $\ket{s}\in \{\ket{+}, \ket{-}\}$ is an eigenvector of $\hat{H}_\mathrm{m}$. Changing from the basis $\{\ket{\mathrm{e}}, \ket{\mathrm{g}}\}$ to the basis $\{\ket{L}, \ket{R}\}$ is achieved by the unitary matrix
\begin{subequations}
\begin{align}
    V &= \frac{1}{\sqrt{1+a^2}} \begin{pmatrix}
        1 & -a\\
        a & 1
    \end{pmatrix}, \text{ with}\\
    a &= \frac{\Omega - \epsilon}{t},\\
    \Omega &= \sqrt{\epsilon^2 + t^2}.
\end{align}
\end{subequations}
The interaction Hamiltonian in the eigenbasis of $\hat{H}_0$ is given by
\begin{equation}
    H_\mathrm{int}^\mathrm{c} = -\delta\gamma (V^\dagger\sigma_R V) \otimes \sigma_z = \frac{-\delta\gamma}{1+a^2}\begin{pmatrix}
        a^2 & a\\
        a & 1
    \end{pmatrix}\otimes\begin{pmatrix}
        1 & 0\\
        0 & -1
    \end{pmatrix}.
    \label{eq:interaction-hamiltonian}
\end{equation}
Here, $\sigma_R$ denotes the matrix of $\ket{R}\bra{R}$ in the basis $\{\ket{L},\ket{R}\}$.
The diagonal matrix elements of $H_\mathrm{int}^\mathrm{c}$ are the first-order energy corrections.

Now we show how the time evolution operator of the total system is calculated up to first order in $\delta\gamma $ using time dependent perturbation theory. The state is expanded in the eigenbasis of $\hat{H}_0$:
\begin{equation}
\begin{split}
    \ket{\psi(\tau)} &= \sum_i \sum_{s\in \{+, -\}} c_{i, s}(\tau) e^{-\frac{i}{\hbar}E_{i,s}\tau}\ket{e_i, s}\\
    &= \sum_i \sum_{s\in \{+, -\}} \left(c_{i, s}^{(0)}(\tau)+c_{i, s}^{(1)}(\tau)\right) e^{-\frac{i}{\hbar}E_{i,s}\tau}\ket{e_i, s} \\
    & +\order{\delta\gamma^2}.
\end{split}
\end{equation}
Here $E_{i, s}=E_{i}+E_{s}$ is an eigenenergy of $\hat{H}_0$; and $E_i$ and $E_s$ are eigenenergies of the $\hat{H}_\mathrm{charge}$ and $\hat{H}_\mathrm{m}$ respectively. At the second equation mark, we split the expansion coefficients $c_{i,s}(\tau)$ into zeroth and first order terms $c_{i, s}^{(0)}$ and $c_{i, s}^{(1)}$. According to first order time dependent perturbation theory, we have
\begin{subequations}
\begin{align}
\begin{split}
    &c_{i, s}^{(1)}(\Delta\tau)= \\
    &= -\frac{i}{\hbar}\sum_{j,s'}\left[\int_0^{\Delta\tau} d\tau\bra{i,s}\hat{H}_\mathrm{int}^{\mathrm{c}}\ket{j,s'} \cdot  e^{-\frac{i}{\hbar}(E_{j,s'}-E_{i, s})\tau} c_{j,s'}^{(0)}\right]\\
    &=  -\frac{i}{\hbar}\sum_{j} s (H_\mathrm{int}^\mathrm{q})_{i,j} c_{j,s}^{(0)} \int_0^{\Delta\tau} dt'e^{-i\omega_{i,j}\tau}\\
    &= \sum_j s C_{ij}(\Delta\tau)c_{j,s}^{(0)}, \quad \text{where}
\end{split}\\
\begin{split}
    &C_{ij}(\Delta\tau)=-\frac{i}{\hbar}\delta_{i,j} E_j^\mathrm{int}\Delta\tau\\
    &  \qquad \qquad \quad  +(1-\delta_{i,j})\frac{(H_\mathrm{int}^\mathrm{q})_{i,j}}{\hbar \omega_{i,j}} \left(e^{-i\omega_{i,j}\Delta\tau}-1\right).
\end{split}
\label{eq:coeffs-first-order}
\end{align}
\end{subequations}
We used the notations $\omega_{i,j}= (E_i-E_j)/\hbar$ and $E_j^\mathrm{int}=(H_\mathrm{int}^\mathrm{q})_{jj}.$ At the second equation mark, we used that the interaction Hamiltonian has the form $\hat{H}_\mathrm{int}^{\mathrm{c}} = \hat{H}_\mathrm{int}^\mathrm{q}\otimes \hat{s}_x$. For the charge qubit, $H_\mathrm{int}^\mathrm{q}$ is the first factor of the tensor product in Eq.~\eqref{eq:interaction-hamiltonian}. Therefore, we can write $\bra{i,s}\hat{H}_\mathrm{int}^{\mathrm{c}}\ket{j,s'}=s(H_\mathrm{int}^\mathrm{q})_{i,j} \delta_{s,s'}$. \par
The time evolution operator $\hat{U}(\tau)$ is related to the expansion coefficients as follows:
\begin{equation}
    \ket{\psi(\tau)}=\hat{U}_\mathrm{tot}\ket{\psi(0)} = \sum_{i, s} c_{i,s}(\tau)e^{-\frac{i}{\hbar}E_{i,s}\Delta\tau}\ket{e_i, s}.
\end{equation}
Therefore, up to first order, using Eq.~\eqref{eq:coeffs-first-order} we get 
\begin{equation}
    (U_\mathrm{tot})_{(i,s),(j,s')} = \delta_{s,s'}(\delta_{ij}+sC_{ij}(\Delta\tau))e^{-\frac{i}{\hbar}E_{i,s}\Delta\tau}.
    \label{eq:time-evol-op}
\end{equation}
Using Eqs.~\eqref{eq:time-evol-op}, \eqref{eq:meas-ops-def}, we get the matrix form of the measurement operators up to $\order{\delta\gamma}$ in the unperturbed basis:
\begin{subequations}
\begin{align}
\begin{split}
    (M_0)_{ij} &=\delta_{ij} e^{-\frac{i}{\hbar} E_i      \Delta\tau}  \cos\left(\frac{\gamma+E_i^\mathrm{int}}{\hbar}\Delta\tau\right)\\
    &-i(1-\delta_{ij})\frac{\sqrt{2}}{2} 
    \frac{(H_\mathrm{int}^\mathrm{q})_{ij}}{E_j-E_i}\left(e^{-     \frac{i}{\hbar}E_j\Delta\tau}-e^{-\frac{i}  {\hbar}E_i\Delta\tau}\right),
\end{split}\\
\begin{split}
    (M_1)_{ij} &=\delta_{ij} (-i)e^{-\frac{i}{\hbar}E_i \Delta\tau} \sin\left(\frac{\gamma+E_i^\mathrm{int}}{\hbar}\Delta\tau\right) \\
    &+(1-\delta_{ij})\frac{\sqrt{2}}{2}\frac{(H_\mathrm{int}^\mathrm{q})_{ij}}{E_j-E_i}\left(e^{-\frac{i}{\hbar}E_j\Delta\tau}-e^{-\frac{i}{\hbar}E_i\Delta\tau}\right).
\end{split}
\end{align}
\label{eq:meas-ops-general}
\end{subequations}
By substituting the energy values of the charge qubit into Eq.~\eqref{eq:meas-ops-general}, we get
\begin{subequations}
\begin{align}
    &M_0 = \begin{pmatrix}
        c_L e^{-\frac{i}{\hbar}\Omega\Delta\tau}& -\frac{\delta\gamma t}{2\Omega^2} s_L \sin\left(\frac{\Omega\Delta\tau}{\hbar}\right)  \\
        -\frac{\delta\gamma t}{2\Omega^2} s_L \sin\left(\frac{\Omega\Delta\tau}{\hbar}\right)
        & c_R e^{\frac{i}{\hbar}\Omega\Delta\tau}
    \end{pmatrix},\\
    &M_1 = \begin{pmatrix}
        -i s_L e^{-\frac{i}{\hbar}\epsilon\Delta\tau} &  \frac{i\delta\gamma t}{2\Omega^2} c_L \sin\left(\frac{\Omega\Delta\tau}{\hbar}\right)  \\
        \frac{i\delta\gamma t}{2\Omega^2} c_L \sin\left(\frac{\Omega\Delta\tau}{\hbar}\right) & -i s_R e^{\frac{i}{\hbar}\Omega\Delta\tau} 
        \end{pmatrix}, \text{ where}\\
    &c_L = \cos\left[\frac{\Delta\tau}{\hbar}\left(\gamma-\frac{\delta\gamma a^2}{1+a^2}\right)\right], \nonumber\\
    &c_R = \cos\left[\frac{\Delta\tau}{\hbar}\left(\gamma-\frac{\delta\gamma}{1+a^2}\right)\right],\nonumber\\
    &s_L = \sin\left[\frac{\Delta\tau}{\hbar}\left(\gamma-\frac{\delta\gamma a^2}{1+a^2}\right)\right],\nonumber\\
    &s_R = \sin\left[\frac{\Delta\tau}{\hbar}\left(\gamma-\frac{\delta\gamma}{1+a^2}\right)\right].\nonumber
\end{align}
\end{subequations}
Note that these expressions depend on $t$ through $a$ and $\Omega$.

The transfer matrix describing the unconditional time evolution is given by
\begin{equation}
    \Upsilon = \Upsilon_0+\Upsilon_1 = (M_0)^* \otimes M_0 + (M_1)^* \otimes M_1.
\end{equation}
Using Eq.~\eqref{eq:meas-ops-general}, we get
\begin{subequations}
\begin{align}
\begin{split}
    &\Upsilon_{\overline{ik},\overline{jl}} = \delta_{ij}\delta_{kl} e^{-\frac{i}{\hbar}(E_k-E_i)\Delta\tau} + \order{\delta\gamma^2}\\
    &\quad+ (1-\delta_{ij})(1-\delta_{kl})T_{ij}^*T_{kl}+ \order{\delta\gamma^3}\\
    &\quad+ \delta_{ij} (1-\delta_{kl}) \cdot \order{\delta\gamma^2}\\
    &\quad+ (1-\delta_{ij}) \delta_{kl} \cdot \order{\delta\gamma^2},    \text{ where}
\end{split}\\
\begin{split}
        &T_{ij} =  \frac{(H_\mathrm{int}^\mathrm{q})_{ij}}{E_j-E_i}\left(e^{-\frac{i}{\hbar}E_j\Delta\tau}-e^{-\frac{i}{\hbar}E_i\Delta\tau}\right).
\end{split}
\end{align}
\end{subequations}
Recall that we use the notation $\overline{ik} = i\cdot d + k $. We got this result form first order time dependent perturbation theory, so the result is accurate to first order in $\delta\gamma/\epsilon$. However, some matrix elements are accurate to second order. We have $\Upsilon_{\overline{ij},\overline{kl}}= (M_0)_{ij}^*  (M_0)_{kl} + (M_1)_{ij}^*  (M_1)_{kl}$, therefore, if the matrix elements $(M_{0/1})_{ij}$ and $(M_{0/1})_{kl}$ do not contain zeroth order terms, the resulting matrix element $\Upsilon_{\overline{ij},\overline{kl}}$ of the transfer matrix is accurate to second order.  

For the charge qubit, we get
\begin{equation}
\begin{split}
    \Upsilon &= \begin{pmatrix}
        1-b & \order{\delta\gamma^2} & \order{\delta\gamma^2} & b \\
        \order{\delta\gamma^2} & e^{-\frac{i}{\hbar}2\Omega \Delta\tau} & \order{\delta\gamma^2} & \order{\delta\gamma^2} \\
        \order{\delta\gamma^2} & \order{\delta\gamma^2} & e^{\frac{i}{\hbar}2\Omega \Delta\tau} & \order{\delta\gamma^2} \\
        b & \order{\delta\gamma^2} & \order{\delta\gamma^2} & 1-b 
    \end{pmatrix}, \text{ where}\\
    b &= \frac{a^2}{(1+a^2)^2}\frac{\delta\gamma^2}{(2\epsilon)^2} 4\sin^2\left(\frac{\Omega\Delta\tau}{\hbar}\right) \approx \frac{t^2\delta\gamma^2}{4\epsilon^4} \sin^2\left(\frac{\epsilon\Delta\tau}{\hbar}\right).
\end{split}
\end{equation}
At the final equation mark, we rewrote $b$ to leading order in $t/\epsilon$. This is a relevant limit, since experimentally $t\ll \epsilon $ usually holds during readout.
The $-b$ term in the matrix element $\Upsilon_{00}$ is not obtained directly from the first order calculation, however, $\Upsilon_{00}+\Upsilon_{30}=1$, since $\Upsilon$ represents a trace preserving map. Using this relation, we can calculate $\Upsilon_{00}$ up to $\order{\delta\gamma^2}$, since we know $\Upsilon_{03}$ up to $\order{\delta\gamma^2}$.  

We aim to diagonalize the transfer matrix $\Upsilon$ to extract the relaxation rate from it. It is diagonal up to $\order{\delta\gamma^2}$, but the 0th and 3rd diagonal elements are the same, therefore, the offdiagonal matrix elements $\Upsilon_{03}$ and $\Upsilon_{30}$ give $\order{\delta\gamma^2}$ contributions to the eigenvalues. For the relaxation rate, the relevant eigenmode of $\Upsilon$ is $(1, 0, 0, -1)$ and the corresponding eigenvalue is $1-2b$. Therefore, we get
\begin{equation}
    \rho_{ee}-\rho_{gg} \sim e^{-\Gamma_\mathrm{rel}^\mathrm{c}\tau},
\end{equation}
where we defined the relaxation rate
\begin{equation}
    \Gamma_\mathrm{rel}^\mathrm{c} = \frac{2b}{\Delta\tau} = \frac{1}{2}\frac{t^2\delta\gamma^2}{\epsilon^4 \Delta\tau}\sin^2\left(\frac{\epsilon\Delta\tau}{\hbar}\right).
\end{equation}
This expression is true up to second order in both $\delta\gamma$ and $t$.

\begin{figure}
    \centering
    \includegraphics[width=\linewidth]{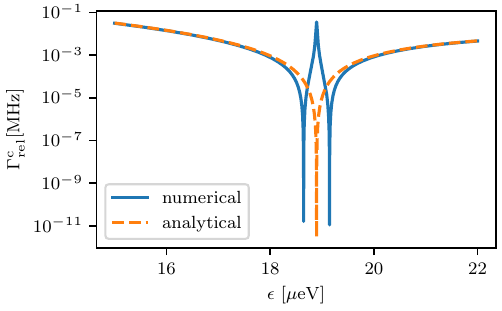}
    \caption{The relaxation rate as a function of the detuning. Analytical results match the numerics well except for the regions $2\Omega \Delta\tau/\hbar \approx 2k\pi$, $k\in\mathbb{Z}$.}
    \label{fig:rel_rate-detuning}
\end{figure}

It is important to note that our expression is only valid if $\abs{e^{\pm \frac{i}{\hbar}2\Omega\Delta\tau}-1}\gg \frac{\delta\gamma}{\epsilon}$, since otherwise other offdiagonal elements of $\Upsilon$ can give relevant contributions to the eigenvalues as well. Therefore, the expression should not be used, if $2\Omega\Delta\tau/\hbar\approx 2k\pi$, $k\in\mathbb{Z}$. This is confirmed by Fig.~\ref{fig:rel_rate-detuning}, where both analytically and numerically calculated values of the relaxation rate are shown. For the experimentally relevant parameter set that we used in our numerical calculations, the analytical results match the numerical data well as shown by the inset of Fig.~2(b).

For the spin qubit, in the presence of QPC-induced Zeeman field modulation described by Eq.~(8), the same calculation shows that transitions between the states $\ket{\downarrow\downarrow}$ and $\ket{\downarrow\uparrow}$; and the states $\ket{\uparrow\uparrow}$ and $\ket{\uparrow\downarrow}$ are possible, leading to
\begin{subequations}
\begin{align}
    &\rho_{\downarrow\downarrow, \downarrow\downarrow} - \rho_{\downarrow\uparrow, \downarrow\uparrow} \sim e^{-\Gamma_\mathrm{leak}\tau},\\
    &\rho_{\uparrow\uparrow, \uparrow\uparrow} - \rho_{\uparrow\downarrow, \uparrow\downarrow} \sim e^{-\Gamma_\mathrm{leak}\tau}.
\end{align}
\end{subequations}
The appearing leakage rate is given by Eq.~(10).
It characterizes leakage from the computational state $\ket{\downarrow\downarrow}$ to the state $\ket{\downarrow\uparrow}$, which is not in the computational subspace during readout.
Similarly to the charge qubit, this result is not valid if $Z_R \Delta\tau/\hbar\approx 2k\pi$, $k\in\mathbb{Z}$. For the numerically investigated parameter values, the analytical formula gives a good estimate of the relaxation rate as shown by the inset of Fig.~3g.

\section{Analytical formula for readout fidelity and ideal integration time}
\label{sec:id-integration-time}
Here, we derive an analytical estimate for the fidelity. We obtain an order of magnitude estimate for the ideal integration time by finding the minimum of this infidelity function.

The estimate of the infidelity is
\begin{equation}
\begin{split}
    1-\mathcal{F} &\approx 1-\Phi\left(\sqrt{2\Gamma_\mathrm{m} \tau_\mathrm{int}}\right) \\
    &+ \frac{1}{2}\left(1-e^{-\frac{\Gamma_\mathrm{rel}}{2} \tau_\mathrm{int} }I_0\left(\frac{\Gamma_\mathrm{rel}}{2}\tau_\mathrm{int}\right)\right),
\end{split}
\label{eq:infidel-est}
\end{equation}
where $I_0$ is the modified Bessel function of the first kind and $\Phi$ is the cdf of the normal distribution. This analytical estimate of the fidelity is plotted in Fig.~2a together with the numerical results. We now discuss the considerations that lead to this estimate. 
The magnitude of the infidelity is determined by different features for short and long times. For very short times, for which relaxation is negligible, the current distributions corresponding to the two computational states are well approximated by Gaussians as shown by Fig.~\ref{fig:bloch-spheres}a,b. The infidelity is then due to the overlap of these Gaussians, which gives the contribution $1-\Phi\left(\sqrt{2\Gamma_\mathrm{m} \tau_\mathrm{int}}\right)$.

\begin{figure}
    \centering
    \includegraphics[width=\linewidth]{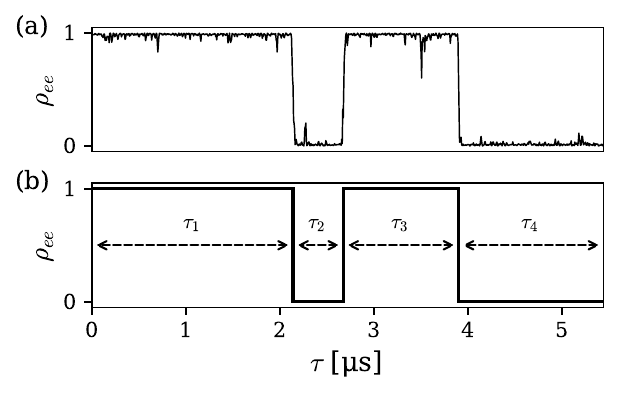}
    \caption{Occupation probability of the excited state for a typical trajectory. (a) Trajectory generated numerically from the Qubit Measures Qubit model with the same parameters as Fig.~\ref{fig:bloch-spheres}. (b) Possible trajectory in a telegraph process that approximates the trajectory coming from the Qubit Measures Qubit model well. Waiting times between jumps are marked on the figure.}
    \label{fig:trajectories}
\end{figure}

On the other hand, for long times the infidelity is mostly due to relaxation. 
Our estimate of this component of the infidelity is based on the observation that the qubit spends most of the time close to one of the computational basis states (see Fig~\ref{fig:trajectories}). The expectation value of the current is $\Bar{I}_{\mathrm{e}}$ ($\Bar{I}_{\mathrm{g}}$) when the qubit is in state e (g). The time average of the current for a given trajectory is then well approximated by
\begin{equation}
    \langle I\rangle = \frac{\tau_{\mathrm{e}}}{\tau_\mathrm{int}}\bar{I}_{\mathrm{e}} + \frac{\tau_{\mathrm{g}}}{\tau_\mathrm{int}}\bar{I}_{\mathrm{g}}
\end{equation}
for long time $\tau_\mathrm{int}$, where $\tau_{\mathrm{e}}$ and $\tau_{\mathrm{g}}$ are the times spent in the excited and ground states respectively. The outcome e is inferred if $\abs{\Bar{I}-\bar{I}_{\mathrm{e}}} < \abs{\bar{I}-\bar{I}_{\mathrm{g}}}$. Therefore, when starting from the initial state $\ket{\mathrm{g}}$, an inference error is made if $\tau_{\mathrm{e}} > \tau_{\mathrm{g}}$, which happens with probability $P(\tau_{\mathrm{e}}>\tau_{\mathrm{g}}|\ket{\mathrm{g}})$. Similarly, for the initial state $\ket{\mathrm{e}}$, an inference error is made if $\tau_{\mathrm{g}} > \tau_{\mathrm{e}}$, which happens with probability $P(\tau_{\mathrm{g}}>\tau_{\mathrm{e}}|\ket{\mathrm{e}})$. 
The rate of jumps from $\ket{\mathrm{e}}$ to $\ket{\mathrm{g}}$ is the same as the rate of jumps from $\ket{\mathrm{g}}$ to $\ket{\mathrm{e}}$, therefore, the probabilities $P(\tau_{\mathrm{e}}>\tau_{\mathrm{g}}|\ket{\mathrm{g}})$ and $P(\tau_{\mathrm{g}}>\tau_{\mathrm{e}}|\ket{\mathrm{e}})$ are the same. For long times this probability gives the infidelity (see Eq.~(2)). Therefore, the second error term in the infidelity estimate Eq.~\eqref{eq:infidel-est} is $P(\tau_{\mathrm{g}}>\tau_{\mathrm{e}}|\ket{\mathrm{e}})=P(\tau_{\mathrm{e}}>\tau_{\mathrm{g}}|\ket{\mathrm{g}})$.

Now we calculate the probability $P(\tau_{\mathrm{g}}>\tau_{\mathrm{e}}|\ket{\mathrm{e}})$. To do this, we assume that the state is in exactly one of the computational basis states during the whole measurement and jumps between the two states are instantaneous and independent, as shown in Fig.~\ref{fig:trajectories}b. The rate of jumps from one state to the other is $\Gamma_\mathrm{rel}/2$ (see Eq.~(4)). Therefore, the problem is described by a symmetric random telegraph process. The jumps of a symmetric random telegraph process occur according to a Poisson process. We denote the waiting times between jumps as $\tau_1$, $\tau_2$, ..., $\tau_i$, ... (see Fig.~\ref{fig:trajectories}b).

Now we assume that exactly $n$ jumps occur during the integration time $\tau_\mathrm{int}$, which split the integration time into $n+1$ waiting times. Then the conditional probability of an inference error given that $n$ jumps occurred is
\begin{equation}
\begin{split}
        &P(\tau_{\mathrm{g}}>\tau_{\mathrm{e}}|\ket{\mathrm{e}}, n) =\\ 
        &P\left(\tau_1+\tau_3+...+\tau_{2\left\lfloor \frac{n}{2}\right\rfloor +1} < \tau_2+\tau_4+...+\tau_{2\left\lfloor \frac{n+1}{2}\right\rfloor}\right).
\end{split}
\end{equation}
The waiting times are described by the multivariate random variable $\underline{\tau} = (\tau_1, ..., \tau_{n+1})$. 

The probability density function $f_{\underline{\tau}}$ of $\underline{\tau}$ is symmetric under exchanging the different waiting times, i.e, $f_{\underline{\tau}}(..., \tau_i, ..., \tau_j, ...) = f_{\underline{\tau}}(..., \tau_j, ..., \tau_i, ...)$ \cite{spacings}. Therefore, the probability $P(\tau_{\mathrm{g}}>\tau_{\mathrm{e}}|\ket{\mathrm{e}}, n)$ is the same as the probability that the sum of the first $\lfloor(n+1)/2\rfloor$ waiting times is less then $\tau_\mathrm{int}/2$, i.e. the probability that at least $\lfloor(n+1)/2\rfloor$ jumps occur in the time interval $[0, \tau_\mathrm{int}/2]$. The jump times are uniformly distributed in the interval $[0, \tau_\mathrm{int}]$ and are independent. Therefore, we get
\begin{equation}
\begin{split}
    &P(\tau_{\mathrm{g}}>\tau_{\mathrm{e}}|\ket{\mathrm{e}}, n) =\\
    &\frac{1}{2^n} \sum_{i=\lfloor n/2\rfloor+1}^{n} {n \choose i} = \begin{cases}
    \frac{1}{2} - \frac{1}{2^{n+1}} {n \choose n/2}, &\text{for even }n\\
    \frac{1}{2}, &\text{for odd }n.
    \end{cases}
\end{split}
\end{equation}
This is the probability of an inference error given that $n$ jumps occur. 

The probability distribution of the number of jumps $n$ is a Poisson distribution with parameter $\lambda= \tau_\mathrm{int}\Gamma_\mathrm{rel}/2 $. Therefore, the probability of an inference error due to relaxation is given by
\begin{equation}
\begin{split}
    &P(\tau_{\mathrm{g}}>\tau_{\mathrm{e}}|\ket{\mathrm{e}}) = \sum_{n=0}^\infty P(\tau_{\mathrm{g}}>\tau_{\mathrm{e}}|\ket{\mathrm{e}},n) P(n) \\
    &=\frac{1}{2} \left(1 - \sum_{i=0}^\infty \left(\frac{\lambda}{2}\right)^{2i}\frac{1}{(i!)^2}e^{-\lambda}\right)= \frac{1}{2}\left(1-e^{-\lambda} I_0(\lambda)\right),
\end{split}
\end{equation}
where $I_0$ is the modified Bessel function of the first kind. This probability is the second term in the estimate of the infidelity Eq.~\eqref{eq:infidel-est}.

Now we derive an order of magnitude estimate for the ideal integration time $\tau_\mathrm{id}$ by finding the minimum of the infidelity estimate. 
For $\tau \ll \tau_\mathrm{id}$ and $\tau \gg \tau_\mathrm{id}$, we expect that Eq.~\eqref{eq:infidel-est} provides a good approximation on the readout infidelity, since in each of these regions only one of the error mechanisms is significant.  
This suggests that the minimum of the infidelity estimate in Eq.~\eqref{eq:infidel-est} provides a good order-of-magnitude estimate of the ideal integration time.

The derivative of Eq.~\eqref{eq:infidel-est} is zero at $\tau_\mathrm{id}$, which leads to
\begin{equation}
    \frac{x}{2} e^{-x} \left( I_{0}(x) - I_{1}(x)\right) - \frac{\sqrt{\Gamma_\mathrm{m} \tau_\mathrm{id}} e^{- \Gamma_\mathrm{m} \tau_\mathrm{id}}}{2 \sqrt{\pi}}=0,
    \label{eq:fidel-der}
\end{equation}
where $x=\Gamma_\mathrm{rel}\tau_\mathrm{id}/2$. For high fidelity measurements, we have $\Gamma_\mathrm{rel}\ll 1/\tau_\mathrm{id} \lesssim \Gamma_\mathrm{m}$. We solve Eq.~\eqref{eq:fidel-der} using this assumption, which leads to
\begin{equation}
    \tau_\mathrm{id} = \frac{1}{\Gamma_\mathrm{m}}\left[\ln\left(\frac{2\Gamma_\mathrm{m}}{\Gamma_\mathrm{rel}}\right)- \frac{1}{2}\ln\left(\Gamma_\mathrm{m} \tau_\mathrm{id} \pi\right)\right].
\end{equation}
To get an order of magnitude estimate of $\tau_\mathrm{id}$, we can neglect the second term, since $\Gamma_\mathrm{rel}\ll 1/\tau_\mathrm{id}$:
\begin{equation}
    \tau_\mathrm{id} \sim \frac{1}{\Gamma_\mathrm{m}} \ln\left(\frac{2\Gamma_\mathrm{m}}{\Gamma_\mathrm{rel}}\right) = \frac{2}{\Gamma_\mathrm{m}}\ln\left(\frac{\sqrt{2}\Delta\tau \epsilon}{\hbar \sin\left(\frac{\epsilon\Delta\tau}{\hbar}\right)}\frac{\epsilon}{t}\right).
\end{equation}
The inset of Fig.~2a shows that this gives a good estimate of the numerically calculated values of the ideal integration time. 
Note that the ideal integration time scales logarithmically with the residual tunneling amplitude $t$.

\section{Back-action due to residual tunneling for spin qubit}
\label{sec:spin-residual-tunneling}
We have shown for a charge qubit that a nonzero residual tunneling during charge sensing leads to relaxation due to $[\hat{H}_\mathrm{charge}, \hat{H}_\mathrm{int}^\mathrm{c}]\neq 0$. 
The analogous relation $[\hat{H}_\mathrm{spin}, \hat{H}_\mathrm{int}^\mathrm{s}]\neq 0$ is true for the spin qubit in the presence of a nonzero residual tunneling. 
Therefore, readout induces incoherent transitions between the energy eigenstates of the spin qubit.

The rates of these transitions can be calculated in the way as for the charge qubit (see SM~\ref{sec:rel-rate}). This calculation is performed in the eigenbasis of the spin qubit, therefore, the interaction Hamiltonian in Eq.~(7) needs to be rewritten in the eigenbasis of $\hat{H}_0 = \hat{H}_{\mathrm{spin}}+ \hat{H}_\mathrm{m}$, which is $\left\{\ket{\Tilde{S}_{(2,0)}}, \ket{\Tilde{S}_{(0,2)}}, \ket{\Tilde{\uparrow\downarrow}}, \ket{{\uparrow\uparrow}}, \ket{{\downarrow\downarrow}}, \ket{\Tilde{\downarrow\uparrow}} \right\}$. 
We tilded the states to indicate that due to the finite tunneling amplitude, the energy eigenstates are not exactly the eigenstates of the spin operators. 

Up to first order in $t$, we obtain 
\begin{subequations}
\begin{align}
    H_\mathrm{int}^\mathrm{s} &= -\delta\gamma\begin{pmatrix}
    0 & 0 & 0 & 0& 0 & 0\\
    0 & 1 & a_1 & 0& 0 & a_2\\
    0 & a_1 & 0 & 0& 0 & 0\\
    0 & 0 & 0 & 0& 0 & 0\\
    0 & 0 & 0 & 0& 0 & 0\\
    0 & a_2 & 0 & 0& 0 & 0\\
    \end{pmatrix}\otimes
    \begin{pmatrix}
        1 & 0\\
        0 & -1
    \end{pmatrix},\\
    a_1 &= \frac{t}{\epsilon-U+\Delta Z},\\
    a_2 &= -\frac{t}{\epsilon-U-\Delta Z},
\end{align}
\end{subequations}
where $\Delta Z = (Z_L-Z_R)/2$.
\begin{figure}
    \centering
    \includegraphics[width=\linewidth]{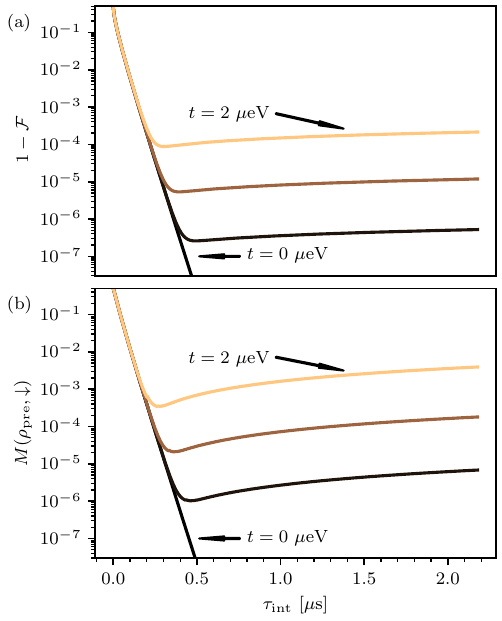}
    \caption{Measurement benchmarks for the spin qubit with nonzero residual tunneling. (a) Infidelity and (b) mixedness for $\rho_{\mathrm{pre}} = \left(\ket{\Tilde{S}_{(0,2)}}\bra{\Tilde{S}_{(0,2)}}+\ket{\downarrow\downarrow}\bra{\downarrow\downarrow}\right)/2$ as a function of integration time.
    Both quantities are calculated numerically for the parameters $U=1$~meV $\epsilon=U+40$~$\mathrm{\mu eV}$, $Z_L = 11$~$\mathrm{\mu eV}$, $Z_R=9$~$\mathrm{\mu eV}$, $\gamma=5$~$\mathrm{\mu eV}$, $\delta\gamma=0.5$~$\mathrm{\mu eV}$. The values of the residual tunneling for the different curves are $t=2$~$\mathrm{\mu eV}$, $t=0.5$~$\mathrm{\mu eV}$, $t=0.1$~$\mathrm{\mu eV}$ and $t=0$~$\mathrm{\mu eV}$.}
    \label{fig:quantities-spin}
\end{figure}
An analogous derivation to the one in SM~\ref{sec:rel-rate} then gives the transition rates
\begin{subequations}
\begin{align}
    &b_{\ket{\Tilde{S}_{(0, 2)}}\leftrightarrow \ket{\Tilde{\uparrow \downarrow}}} = \frac{4\delta\gamma^2 t^2}{(\epsilon-U+\Delta Z)^4} \sin^2\left(\frac{\epsilon-U+\Delta Z}{2\hbar} \Delta\tau\right), \\
    &b_{\ket{\Tilde{S}_{(0, 2)}}\leftrightarrow \ket{\Tilde{\downarrow \uparrow}}} = \frac{4\delta\gamma^2 t^2}{(\epsilon-U-\Delta Z)^4} \sin^2\left(\frac{\epsilon-U-\Delta Z}{2\hbar} \Delta\tau\right), 
\end{align}
\end{subequations}
which are accurate to second order in $\delta\gamma \cdot t/(\epsilon-U\pm \Delta Z)^2$. These rates are analogous to the transition rate of the charge qubit: incoherent transitions are induced between the eigenstates of the data qubit Hamiltonian that are connected by finite matrix elements of the interaction Hamiltonian.

The measurement benchmarks infidelity and mixedness as a function of integration time show the same qualitative behavior as for the charge qubit as shown by the numerical results in Fig.~\ref{fig:quantities-spin}. The same analytical estimate for the fidelity is not possible for the spin qubit as for the charge qubit (see SM~\ref{sec:id-integration-time}), since transitions happen between three states, not only two.

\section{Comparison with the stochastic Schrödinger equation}
\label{sec:sde}
Here, we compare the QMQ model with a previously used description of QPC-based readout that uses a stochastic Schrödinger equation (SSE) or a stochastic master equation used in Refs.~\onlinecite{Goan_2001_1, Goan_2001_2}. We make the comparison using the example of charge qubit readout, however, our statements straightforwardly generalize to spin qubit readout. We show that the SSE description and the QMQ model give the same qualitative predictions. We also make quantitative comparison by comparing the three characteristic rates of the dynamics: the measurement rate, the dephasing rate and the measurement induced relaxation rate. We find that for the measurement and dephasing rates there is a factor of 2 difference between the predictions, and argue that the prediction of the QMQ model is the correct one. For the relaxation rate, the difference is a factor of around 1.38 for the parameter set used in our numerical calculations (see SM~\ref{sec:exp-params}).

It is an open question which model gives the better quantitative predictions. Both models have assumptions that might not be realistic. The SSE description assumes that transmission through the QPC is instantaneous; and there is a finite waiting time between electrons reaching the QPC. Conversely, the QMQ model assumes a finite transmission time; and that the next electron reaches the QPC instantaneously after the previous electron is transmitted or reflected. It is reassuring that despite the differing assumptions the qualitative predictions of the two descriptions are the same and quantitative differences are relatively small.

Now we briefly introduce the SSE description, using the notation of our paper where possible. The electron tunneling rates of the QPC are given by $D = \abs{\mathcal{T}}^2$ and $D' = \abs{\mathcal{T}+\chi}^2$ given that the qubit is in state $\ket{L}$ and $\ket{R}$ respectively, where $\chi$ and $\mathcal{T}$ are complex tunneling amplitudes. For more details, see Ref.~\onlinecite{Goan_2001_1}. For simplicity, we assume both $\mathcal{T}$ and $\chi$ to be real. The conditional time evolution is then given by 
\begin{equation}
\begin{split}
    d\ket{\psi_c (\tau)} &= 
    \left[dN_\mathrm{tr}(\tau) \left(\frac{\mathcal{T}+\chi \hat{n}_R}{\sqrt{\mathcal{P}_\mathrm{tr}(\tau)}}-1\right) \right.\\
    -d\tau &\left.\left(\frac{i}{\hbar}\hat{H}_\mathrm{charge}+ \frac{\abs{\mathcal{T}+\chi \hat{n}_R}^2}{2} - \frac{\mathcal{P}_\mathrm{tr}(\tau)}{2} \right) \right]\ket{\psi_c(\tau)},\\
    \mathcal{P}_\mathrm{tr}(\tau) &= D + (D'-D)\langle \hat{n}_R \rangle_\mathrm{\psi_c(\tau)},
\end{split}
\label{eq:sde-normalized}
\end{equation}
where $\hat{n}_R$ is the number operator of the right dot. The number of transmitted electrons $dN_\mathrm{tr}(\tau)$ between times $\tau$ and $\tau + d\tau$ is either 0 or 1, furthermore, $\mathbb{E}[dN_\mathrm{tr}(\tau)] = \mathcal{P}_\mathrm{tr}(\tau)d\tau$. This equation can be used to generate quantum trajectories together with the corresponding current.

We can match the parameters of the SSE description with the parameters of the QMQ model by requiring that the average currents are the same in the two descriptions. The resulting relations are
\begin{subequations}
\begin{align}
    D &= \frac{p_L}{\Delta \tau} \approx \frac{1}{2}\left(\frac{1}{\Delta\tau} + \frac{\delta\gamma}{\hbar} \right),\\
    D' &= \frac{p_R}{\Delta \tau} \approx \frac{1}{2}\left(\frac{1}{\Delta\tau} - \frac{\delta\gamma}{\hbar} \right),
    \label{eq:param-matching}
\end{align}
\end{subequations}
where we made a first order approximation in $\delta\gamma/\gamma$, which is a good approximation if the modulation of the current by the qubit state is much smaller than the average current. In case of the SSE description, the same condition translates to $\abs{\chi}\ll\mathcal{T}$. Making this assumption, we obtain 
\begin{equation}
    \chi \approx \frac{D'-D}{2\mathcal{T}}\approx -\frac{\delta\gamma}{2\hbar}\sqrt{2\Delta\tau}.
\end{equation}
This equation enables us to compare the predictions of the QMQ model and the SSE description.

In the quantum non-demolition case ($t=0$), the computational basis states $\ket{L}$ and $\ket{R}$ are stationary states of the dynamics. Therefore, the distribution of transmitted electrons $N_\mathrm{tr}$ is given by a Poisson processes with parameters $D$ and $D'$ for the states $\ket{L}$ and $\ket{R}$ respectively. These Poisson distributions can be approximated by normal distributions for $\tau\gg 1/D, 1/D'$. The expectations value of $N_\mathrm{tr}$ is $D\tau$ ($D'\tau$) for the initial state $\ket{L}$ ($\ket{R}$), which is guaranteed to be the same as for the QMQ model by Eq.~\eqref{eq:param-matching}. The variances are also $D\tau$ and $D'\tau$ respectively, which are roughly $\tau/(2\Delta\tau)$ when expressed by the parameters of the QMQ model with the help of Eq.~\eqref{eq:param-matching}. These variances are roughly twice of what the QMQ model predicts. 

Now we compare the predictions of the two models for the three characteristic rates of readout: the measurement rate, the dephasing rate, and the relaxation rate. The measurement rate is determined by approximating the distribution of $N_\mathrm{tr}$ by normal distributions for the states $\ket{L}$ and $\ket{R}$, then calculating the overlap of the two distributions. Straightforward calculation results in
\begin{equation}
    \Gamma_m = \frac{\chi^2}{2} = \frac{1}{4} \left(\frac{\delta\gamma}{\hbar}\right)^2\Delta\tau
\end{equation}
up to first order in $\chi/\mathcal{T}$. This is the half of the prediction of the QMQ model (see Eq.~\eqref{eq:meas-rate-val}). The dephasing rate predicted by the SSE description is also $\Gamma_\mathrm{d} = \chi^2/2$ (see Ref.~\onlinecite{Goan_2001_1}), which is once again half the prediction of the QMQ model.

The measurement rate is directly related to the noise properties of the sensor, which determines the width (standard deviation) of the distribution of the transmitted charge. The standard quantity to characterize the noise properties of mesoscopic conductors is the Fano factor \cite{BLANTER2001}, which is the ratio of the shot-noise of the conductor and Poissonian shot-noise corresponding to the same average current. The Landauer-Büttiker formalism \cite{Landauer, PhysRevLett.57.1761} predicts that the Fano factor of a single-channel QPC with transmission probability $T$ is $F=1-T$ \cite{BLANTER2001}, which is well established by experiments \cite{PhysRevLett.75.3340, PhysRevB.93.195411}. Therefore, for such a QPC tuned to $T=0.5$, the Fano factor is $F=0.5$, which is reproduced by the QMQ model. On the other hand, the SSE description assumes that electrons are transmitted according to a Poisson process, which corresponds to $F=1$. This difference is directly responsible for the difference in the measurement rate.

Finally, we compare the predicted relaxation rates. For the SSE model, it is \cite{Goan_2001_1}
\begin{equation}
\begin{split}
    \Gamma_\mathrm{rel}^\mathrm{c} &= \frac{4t^2 \Gamma_d}{\hbar^2\Gamma_d^2+(2\epsilon)^2}= \frac{1}{\hbar^2} \frac{t^2 \delta\gamma^2}{\delta\gamma^4\Delta \tau^2/\hbar^2+ 4\epsilon^2}\Delta\tau.\\
    &\approx  \frac{t^2 \delta\gamma^2}{4\hbar^2\epsilon^2}\Delta\tau.
    \label{eq:sde-rel-rate}
\end{split}
\end{equation}
At the final equation mark, we used $\delta\gamma\ll \epsilon, \gamma$, to drop the first term in the denominator. For the parameters used in our numerical calculations (see SM~\ref{sec:exp-params}), this gives a factor of around $1.38$ larger relaxation rate then the prediction of the QMQ model (see Eq.~(4)).

\section{Measuring back-action induced leakage}
\label{sec:leakage_detection}
Here, we propose a simple procedure to experimentally measure the infidelity and leakage. 
The procedure is based on some simple assumptions about the dispersion relation of the two-spin states and the possible error sources during readout. Similar procedures should be possible for different or more complicated error models as well.

The first assumption is that the dispersion relations for $t=0$ and $t\neq 0$ look like the ones shown in Fig.~3c. As a result, during spin-to-charge conversion, the following adiabatic state transfers happen: $\ket{\uparrow\uparrow} \rightarrow \ket{\uparrow\uparrow}$, $\ket{\downarrow\downarrow} \rightarrow \ket{\downarrow\downarrow}$, $\ket{\uparrow \downarrow} \rightarrow \ket{\downarrow\uparrow}$, $\ket{\downarrow\uparrow} \rightarrow \ket{S_{(0, 2)}}$. During charge-to-spin conversion, the same state transfers happen in the opposite direction and we assume all state transfers to be perfect. Secondly, we assume the following initialization error: Instead of the state $\ket{\downarrow\uparrow}$, the mixture $\hat{\rho}(\downarrow\uparrow) = (1-q_1-q_2)\ket{\downarrow\uparrow}\bra{\downarrow\uparrow}+q_1\ket{\downarrow\downarrow}\bra{\downarrow\downarrow}+q_2\ket{\uparrow\downarrow}\bra{\uparrow\downarrow}$ is prepared. Other spin states are prepared by applying coherent rotations to the prepared state $\rho(\downarrow\uparrow)$. Thirdly, inference errors might occur during charge sensing. We infer the charge configuration $(0, 2)$ ($(1, 1)$) instead of $(1, 1)$ ($(0, 2)$), with probability $\epsilon_{\uparrow}$ ($\epsilon_{\downarrow}$). Finally, leakage occurs during charge sensing from state $\ket{\downarrow\downarrow}$ to state $\ket{\downarrow\uparrow}$ with probability $\mathcal{L}$. 

\begin{figure*}
    \centering
    \includegraphics[width=\linewidth]{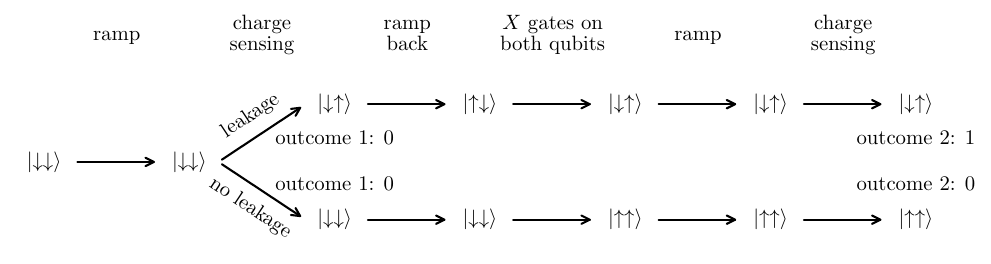}
    \caption{Leakage-detection experiment. The measurement probabilities of different outcomes in the above experiment depend on the probability of leakage (see Eq.~\eqref{eq:leakage-det}). Consequently, leakage can be estimated by performing this experiment many times.}
    \label{fig:leakage-detection}
\end{figure*}

Now, we propose a leakage-detection experiment, which we first present by assuming that all other error probabilities are zero. Then, we list a set of experiments, including the leakage detection experiment, that can be used to estimate the error probabilities $q_1$, $q_2$, $\epsilon_1$, $\epsilon_2$ and $\mathcal{L}$ self-consistently. The leakage-detection experiment is summarized in Fig.~\ref{fig:leakage-detection}. First, the state $\ket{\downarrow\downarrow}$ is prepared and spin-to-charge conversion is performed. Then the charge configuration is measured during which leakage is induced by the measurement with probability $\mathcal{L}$. The outcome of this charge measurement is 0, corresponding to the $(1, 1)$ charge sector. Next, charge-to-spin conversion is performed and $X$ gates are applied to both of the spins. The resulting state after the $X$ gates is $\ket{\downarrow\uparrow}$ if leakage occurred and $\ket{\uparrow\uparrow}$ otherwise. Upon measuring these states again using spin-to-charge conversion and charge sensing, the outcome is $1$ if leakage occurred during the first measurement and $0$ otherwise. Therefore, leakage can be detected using this experiment.

Now, we show that this procedure can also be used to estimate the probability of leakage in the presence of the other error mechanisms listed above. The probabilities of the the four possible combinations of outcomes are
\begin{subequations}
\begin{align}
    p_{(0, 1)} &= \mathcal{L}+\epsilon_\uparrow,\\
    p_{(1, 0)} &= \epsilon_\uparrow,\\
    p_{(1, 1)} &= q_1,\\
    p_{(0, 0)} &= 1-p_{(0, 1)}-p_{(1,0)}-p_{(1, 1)}
\end{align}
\label{eq:leakage-det}
\end{subequations}
up to first order in the small error parameters $q_1$, $q_2$, $\epsilon_{\uparrow}$, $\epsilon_{\downarrow}$ and $\mathcal{L}$. Leakage is then given by $\mathcal{L} = p_{(0,1)}-p_{(1,0)}$. Apparently, the error parameters $\epsilon_{\uparrow}$ and $q_1$ can also be estimated using this leakage-detection experiment. In order to estimate the remaining two error parameters $\epsilon_{\downarrow}$ and $q_2$, two additional experiments need to be performed: (1) the state $\hat{\rho}(\downarrow\uparrow)$ is prepared; (2) the state $\hat{\rho}(\uparrow\downarrow)$ is prepared. Then, the prepared states are measured using spin-to-charge conversion and charge sensing. The outcome probabilities of these experiments are
\begin{subequations}
    \begin{align}
    p_0(\downarrow\uparrow) &= \epsilon_\downarrow+q_1+q_2\\
    p_1(\downarrow\uparrow) &= 1-p_0(\downarrow\uparrow)\\
    p_0(\uparrow\downarrow) &= \epsilon_\uparrow+q_2\\
    p_1(\uparrow\downarrow) &= 1-p_0(\uparrow\downarrow).
\end{align}
\label{eq:additional-experiments}
\end{subequations}
Here, the subscripts and arguments indicate the outcomes and the prepared states respectively.
Combining Eqs.~\eqref{eq:leakage-det} and \eqref{eq:additional-experiments}, we have 5 independent equations, which allow us to estimate all of the error parameters.

\section{Discussion}
\label{sec:discussion}

\subsection{Compatibility of readout sweet spot and dephasing sweet spot}

In the main text, we introduced the readout sweet spot, which is a device configuration where the modulation of the Zeeman field caused by the QPC is parallel to the Zeeman field in the right dot, and the magnitude of the modulation is maximal. In multi-qubit devices, the readout sweet spot could be reached by in-situ tuning of the $g$-tensor. However, there are other dephasing and relaxation sweet spots identified earlier \cite{Piot, Sen_2023, Michal_2023, Mauro_2024, PhysRevLett.127.190501, PRXQuantum.2.010348, PhysRevLett.129.247701}, which pose different conditions for the $g$-tensors. In particular, the requirement for the dephasing sweet spot is that the modulation of the Zeeman field caused by charge fluctuations is perpendicular to the Zeeman field. This raises the question how compatible dephasing and readout sweet spots are.

In principle, the readout sweet spot can be fully compatible with the dephasing sweet spot, although this might require special preconditions (including fine tuning of gate-voltages) in practice. For example, if dephasing is dominated by a few nearby charge fluctuators, then it is possible that the operating point of a qubit is a dephasing sweet spot with respect to the fluctuators, and at the same time, it is also a readout sweet spot with respect to the charge noise caused by the QPC electrons. Here, we refrain from speculating about how to engineer this situation in practice.

Going beyond the special case discussed in the previous paragraph, we claim that it is possible to perform qubit operations at the dephasing sweet spot and readout at the readout sweet spot. We propose two simple strategies to achieve this:
\begin{enumerate}
    \item Use different dots for readout and qubit operations. For example, store the reference spin used for readout in the dot closer to the charge sensor. The g-tensor of this reference spin will be sensitive to charge fluctuations caused by the QPC. Tune this dot to the readout sweet spot, since dephasing of the reference spin does not affect the readout; and tune the dot hosting the data qubit to the dephasing sweet spot.
    \item Operate the qubits at the dephasing sweet spot. When readout is performed, after spin-to-charge conversion, tune the dots to the readout sweet spot.
\end{enumerate}
The feasibility of these strategies is supported by previous experiments, where in-situ tuning of $g$-tensors has been demonstrated  \cite{Crippa_2018,Piot,Hendrickx,John, saezmollejo, carballido, BassiArxiv}.

\subsection{Challenges toward exploiting the readout sweet spot}

We also highlight a number of challenges toward exploiting the readout sweet spot as proposed in the main text.

\emph{Uniformity across a quantum-dot array.}
In a large-scale array of quantum dots, perfect uniformity of spin-orbit interaction (and, hence of $g$-tensors) could guarantee that a `global' readout sweet spot can be found by sweeping the orientation of a homogeneous magnetic field.
For germanium hole spin qubits, early experiments revealed highly different g-tensors in different dots of the same array, which works against the global readout sweet spot. 
However, more recent devices show a much higher degree of uniformity \cite{Seidler}. This progress points toward future devices where the uniformity is improved further, potentially making the concept of the global readout sweet spot experimentally feasible in large-scale arrays.

\emph{Local tunability of spin-orbit interaction.}
Even in a non-uniform array of spin qubits subject to a homogeneous external magnetic field, it is desirable to have all dots used for readout tuned to the readout sweet spot. 
The non-uniformity of the array implies that this requires local fine tuning of the spin-orbit effects by local gate voltages.
Even though the the feasibility of this solution is supported by experimental studies where in-site gate-voltage tuning of the $g$-tensors has been demonstrated, \cite{Crippa_2018,Piot,Hendrickx,John, saezmollejo, carballido, BassiArxiv}, we expect that the complexity of this fine tuning will grow as the size of the array is increased.

\emph{Other type of readout errors.} 
In the main text, the concept of the readout sweet spot refers to the absence of a specific readout error mechanism: leakage due to a transversal coupling between the spin in the readout dot and the electric fluctuations created by the QPC electrons.
It is important to note that having a device tuned to this readout sweet spot does not eliminate all readout error mechanisms. 
One such example is the interplay of QPC shot noise and finite measurement time, a mechanism which is included in our model. 
A further error mechanism, which is straightforward to incorporate into our model, is spin flip due to sources other than the QPC charge noise, e.g., due to phonons, charge traps, gate-voltage fluctuations, etc.
Our model detailed in this work provides a convenient framework for readout optimization even in the presence of such additional sources of readout error.

\end{document}